\documentclass{article}
\usepackage[margin=2.5cm]{geometry}
\usepackage[utf8]{inputenc}
\usepackage{amssymb, amsmath}

\usepackage{url, xcolor, hyperref}
\hypersetup{colorlinks, linkcolor={red!70!black}, citecolor={blue!70!black}, urlcolor={red!70!black}}
\usepackage{graphicx}
\usepackage{settings_custom}

\newcommand{\kgr}{{\kappa_{\rm GR}}}
\newcommand{\krl}{{\kappa_{\rm RL}}}
\newcommand{\egs}{{e_{\rm GS}}}
\newcommand{\kcgr}{\kappa^{\rm cross}_{\rm GR}}
\newcommand{\ksgr}{\kappa^*_{\rm GR}}
\newcommand{\kmingr}{\kappa^{\rm min}_{\rm GR}}

\title{Quenches in the Sherrington-Kirkpatrick model}
\author{Vittorio Erba$^1$, Freya Behrens$^1$, Florent Krzakala$^2$, and Lenka Zdeborová$^1$}
\date{\footnotesize
    $^1$Statistical Physics of Computation Laboratory, \\
    $^2$Information, Learning and Physics Laboratory, \\
    École polytechnique fédérale de Lausanne (EPFL)
    CH-1015 Lausanne
}

\begin{document}
    \maketitle

    \begin{abstract}
The Sherrington-Kirkpatrick (SK) model is a prototype of a complex non-convex energy landscape. Dynamical processes evolving on such landscapes and locally aiming to reach minima are generally poorly understood. Here, we study quenches, i.e. dynamics that locally aim to decrease energy.  
We analyse the energy at convergence for two distinct algorithmic classes, single-spin flip and synchronous dynamics, focusing on greedy and reluctant strategies.
We provide precise numerical analysis of the finite size effects and conclude that, perhaps counter-intuitively, the reluctant algorithm is compatible with converging to the ground state energy density, while the greedy strategy is not.
Inspired by the single-spin reluctant and greedy algorithms, we investigate two synchronous time algorithms, the sync-greedy and sync-reluctant algorithms. These synchronous processes can be analysed using dynamical mean field theory (DMFT), and a new backtracking version of DMFT.
Notably, this is the first time the backtracking DMFT is applied to study dynamical convergence properties in fully connected disordered models.
The analysis suggests that the sync-greedy algorithm can also achieve energies compatible with the ground state, and that it undergoes a dynamical phase transition. 
\end{abstract}

\section{Introduction}
    This work aims to contribute to the understanding of out-of-equilibrium dynamical processes in complex systems. 
    The theoretical analysis of dynamical processes in disordered systems is, in general, challenging. The motivation of this paper is to leverage the main idea coming from recent methodological progress, the Backtracking Dynamical Cavity Method (BDCM) \cite{behrens2023backtracking}, where a dynamical process is studied for a number of timesteps backwards from an attractor rather than forward from a random initial condition. So far, this idea and associated method have been applied to systems on sparse random graphs that are locally tree-like \cite{behrens2023backtracking,behrens2023dynamical}. In this paper, the aim is to apply it to study the dynamics of arguably the most prototypical and broadly studied complex system -- the Sherrington-Kirkpatrick (SK) model of a spin glass.
    
    The  SK model is defined by its Hamiltonian (or energy function)
    \begin{equation}\label{skmodel}
        H_J(x) = - \frac{1}{2\sqrt{N}} \sum_{i, j = 1}^N J_{ij} x_i x_j 
        \quad\text{with}\quad
        x \in \{\pm 1\}^N \, ,
    \end{equation}
    where $J_{ij} = J_{ji}$ is a symmetric matrix with i.i.d. standard Gaussian entries and, for concreteness, zero diagonal $J_{ii}=0$. 
    Its equilibrium properties were famously analyzed through the replica method and associated replica symmetry breaking \cite{parisi1979infinite,beyond}. The value of the ground state energy density, i.e. $\egs = \lim_{N\to \infty} \min_x H_J(x) / N = -0.763219...$ with high probability with respect to $J$, is known to a great precision from solving the corresponding fixed point equations \cite{parisi1979infinite,beyond}. 
    In the following we will call the energy density just energy, for simplicity.
    
    The interaction graph in the SK model is complete as opposed to the sparse locally tree-like graphs for which the BDCM has been developed in \cite{behrens2023backtracking}. Thus, we must generalize the method to the complete graph setting. For the forward dynamics, the counterpart of the dynamical cavity method in sparse systems \cite{hatchett2004parallel,neriCavityApproachParallel2009,mimuraParallelDynamicsDisordered2009} is the dynamical mean field theory (DMFT) in fully connected systems \cite{georges1996dynamical}, and in particular its version developed in \cite{opper94}. 
    In \cite{hwang2019number}, the authors successfully studied attractors of synchronous Glauber dynamics in systems with fully connected interaction graphs through a slight modification of Dynamical Mean Field Theory (DMFT). 
    Building on all these works, we present in this paper a new variant of method called backtracking-DMFT to study the convergence properties of synchronous deterministic dynamics for the SK model. 

    \paragraph{Dynamical properties of neural networks and complex systems.} Dynamical properties of disordered fully-connected spin systems have been widely studied in relation to neural networks, and in particular with models of associative memories \cite{hopfield1982neural}, of asymmetric neural networks \cite{opper94} and of multi-agent interactions \cite{garnierbrun2024unlearnable}. For example, and we only cite here selected works from a very large amount of references, \cite{opper94, hwang2019number, singh2003synchronous} study a synchronous Glauber dynamics in the SK model with non-symmetric couplings, \cite{Henkel_1991} studies dynamical properties of a spin glass with pseudo-inverse couplings, a typical choice to study pattern retrieval phenomena, and \cite{Krauth_1987} studies the retrieval of large-margin patterns in an associative memory model. In this case, the focus is not on the energy, which looses most of its meaning when the couplings are not symmetric, but the typology of dynamic attractors towards which the dynamics converges, the speed at which the dynamics decorrelates from the initial conditions, and whether the degree of asymmetry induces novel static or dynamic phase transitions \cite{opper94, Aguirre-Lopez_2022}. 
    Developing a backtracking-DMFT for the SK model (see the previous motivation paragraph) may allow novel insights into these systems' dynamical properties in the future.

    \paragraph{Zero temperature quenches in the SK model.} 
    Let us now discuss and motivate the questions about the dynamics in the SK model that we will be able to study with the backtracking-DMFT method. For the sparse spin-glass and Ising models, the BDCM has been used in \cite{behrens2023backtracking} to study the properties and limiting energy of zero temperature quenches. Dynamical phase transitions were studied using the BDCM  in cellular automata on sparse random graphs \cite{behrens2023dynamical}. In the SK model, the properties and limiting energy of various types of zero-temperature quenches are also questions of great interest, with no definite answer thus far. 

    The study of the behaviour of synchronous dynamics evolving in complex non-convex landscapes is also of interest because these are used in many practical applications. For instance, in machine learning with neural networks, greedy synchronous algorithms such as gradient descent are widely used to attempt to minimize highly non-convex functions. It is clear that these algorithms do not aim to sample the equilibrium measure, yet it is not clear where they end up in the non-convex landscape. 

    Several versions of the zero-temperature dynamics for the SK model have been studied in the literature. In a zero temperature quench only spins with local magnetic field pointing in the opposite direction are flipped. This can happen one spin at a time or in a synchronous way (a finite fraction of all spins at once). 
    In the single-spin update one may choose to flip the spins in different orders, among which natural choices are the so called greedy, random and reluctant dynamics \cite{Parisi1, contucci03}. 
    In all these dynamical rules the spin configuration evolves by single spin flips towards configurations of lower energy, i.e. they are local descent algorithms for the SK Hamiltonian. 
    The random dynamics flips spins at random among those producing any drop in energy. 
    The greedy dynamics consist of flipping at each time the spin producing the largest drop in energy, while the reluctant consists in flipping the spin producing the least drop in energy. 
    These last two are discrete versions of ``steepest" and ``mildest" descent algorithms. 
    Remarkably, the reluctant dynamics achieves much lower energies at convergence than the greedy one \cite{Parisi1}, albeit with slower convergence time. This may seem not in line with common greedy heuristics in optimization.
    Understanding theoretically why reluctant strategies work better, and to what extent this observation holds, is an open question. 
    This is also extremely relevant given that analogs of the greedy dynamics has been widely studied as a generic purpose minimization heuristics in the field of optimization, while, as far as we know, the corresponding generalisation of the reluctant dynamics has been studied only in relation to the SK model and minor variants \cite{contucci05, contucci05-2}.

    The main goal in this paper is to analyze the dynamics with the backtracking-DMFT and this will be possible only for synchronous updates where all spins are updated at every time step following a common rule, e.g. if the product of the spin with their local magnetic field lays in a certain interval. The energy under such update rules does not necessarily need to decrease, but can also increase, 
    so that these algorithms may converge also to limit cycles
    and not only 
    to local minima of the cost functions. 
    Variants of these synchronous zero temperature update rules, 
    inspired by the greedy and reluctant single-spin flip dynamics,
    are the main object of the theoretical analysis. 
    
    \paragraph{Reaching the ground state energy density in the SK model.}
    A broadly studied and natural combinatorial optimization question is whether configurations having the same energy density as the ground state can be found with efficient (i.e. polynomial time) algorithms. A recent breakthrough answered this long-standing question positively and mathematically rigorously thanks to the Incremental Approximate Message Passing (IAMP) algorithm and its analysis \cite{montanari, alaoui2020algorithmic}. The IAMP algorithm is very efficiently implementable, but it is conceptually rather involved as it e.g. explicitly uses the fixed point equations describing the full-step replica symmetry broken solution of the SK model \cite{parisi1979infinite}. It is another question whether the ground state energy density can be reached using conceptually elementary algorithms such as the above discussed zero-temperature quenches. The present numerical and analytical study of these quenches appears to be compatible with the possibility that in the thermodynamic limit the reluctant version finds configurations with the same energy density as the ground state. We will present the collected evidence and propose several related questions/conjectures for future investigation. 
    
    \paragraph{Main results.} 
    With these motivations, in this paper, we focus on the greedy and the reluctant general optimization strategies, and study them both in a single spin flip setting, such as the one studied in the context of zero-temperature SK dynamics \cite{Parisi1}, and in a synchronous update setting, in which a finite fraction of spins may be flipped at each time step, more akin to the dynamical rules studied in the context of neural network dynamics. 
    The main contributions are the following.
    \begin{itemize}
        \item We update the literature's estimate for the convergence properties of the single spin flip greedy and reluctant dynamics by providing new numerical simulations at considerably larger system sizes. Most interestingly, we find that the energy at convergence for the reluctant dynamics, as well as the behaviour of the local field distribution around the origin, are compatible with the theoretical predictions for the ground state energy density of the SK model. We confirm that the greedy and random algorithms do not converge to the ground state and estimate their limiting energy to be $\approx -0.725$ (to be compared to the ground state $\approx -0.763$).
        \item We introduce the backtracking-DMFT method to study the convergence properties of synchronous dynamical processes in a fully connected disordered model. Through the specific example of zero-temperature synchronous quenches in the SK model, we highlight cases for which the dynamics is rapid and where hence the backtracking DMFT provides new results and insight.
        \item We introduce two zero-temperature synchronous quenches inspired by the greedy and reluctant single spin flip dynamics, the sync-greedy and sync-reluctant algorithm, and we study them numerically and analytically through DMFT (both in the classic forward type, and backtracking). We present a characterisation for the sync-greedy algorithm, providing evidence that it may achieve energies compatible with the ground state energy of the SK model and that it undergoes a dynamical phase transition. We also present observations on the behaviour of the sync-reluctant dynamics, which turns out to be slower and consequently more difficult to analyze with DMFT.
    \end{itemize}

We provide the code used to generate the data and to plot the figures at \url{https://github.com/SPOC-group/quenches-sk}.

\section{Single spin-flip dynamics}\label{sec.single}

    To start with, we consider single spin-flip dynamics.
    We define the local magnetic fields as 
    \begin{equation}
        h_i = \frac{1}{\sqrt{N}} \sum_{j=1}^N J_{ij} x_j \, ,
    \end{equation}
    and notice that
    \begin{equation}
        H_J(x) = - \frac{1}{2\sqrt{N}} \sum_{i, j = 1}^N J_{ij} x_i x_j = - \frac{1}{2} \sum_{i=1}^N x_i h_i \, .
    \end{equation}
    We will call a spin $i$ \textit{unsatisfied} if $x_i h_i < 0$, and satisfied otherwise. 
    
  In order to minimize the Hamiltonian, it is natural to consider heuristics that aim at having  $x_i h_i > 0$ for all $i$. These configurations are local minima of the energy, in the sense that any single spin flip increases the energy.
    We study the following single spin-flip algorithms:
    \begin{itemize}
        \item Random/Sequential: at each time step flip one spin at random among those for which $x_i h_i < 0$ holds. A sequential variant is often considered, in which spins are flipped in order of increasing $i$. The two dynamics behave quantitatively comparably even at finite size $N$, and so we discuss them together.
        \item Greedy: at each time step flip the spin $i$ such that $x_i h_i < 0$ and $|h_i|$ is maximum.
        \item Reluctant: at each time step flip the spin $i$ such that $x_i h_i < 0$ and $|h_i|$ is minimum.
    \end{itemize}
    Notice that any single spin-flip dynamics that flips exclusively unsatisfied spins will converge to a fixed point, as it is easy to see that the energy is a monotonic function along any trajectory, and the configuration space has finite cardinality at any fixed size $N$.

    \subsection{Related previous works}
    The above algorithms have been studied numerically in the literature. In particular \cite{Parisi1} observed that the reluctant algorithm performs best among the three, despite the fact that at every steps it achieves the smallest possible improvement to the energy function which may come as a surprise.
    Previous works differ in terms of interpretations of the numerical findings. In particular: 
    \begin{itemize}
       \item Random/Sequential: In \cite{Parisi1} the authors report $e_{\rm sequential} \approx -0.715$, with finite size correction exponent $\alpha = -0.33$. We will provide different estimates based on simulations of larger sizes. In \cite{Parisi1}  they also report that the sequential algorithm finds systematically configurations with better energies than the greedy at all finite sizes considered, which is confirmed in our simulations.
       \item Greedy: It is claimed in \cite[Figure 4 and 5]{contucci03} that the greedy algorithm does not converge to $\egs$, but to an higher energy $e_{\rm greedy} \approx -0.735$, and has a convergence time of $\caO(N)$ spin flips. This is in qualitative agreement 
       with \cite{Parisi1}, which finds $e_{\rm greedy} \approx -0.708$ at convergence, and \cite{Horner07}, which finds $e_{\rm greedy} \approx -0.729$.
       We remark that the averaging protocol of \cite{contucci03} is different from \cite{Parisi1, Horner07}: in \cite{contucci03} they plot the minimum energy reached for fixed $J$ starting from $\caO(N)$ initial conditions for a system of size $N$, while in e.g. \cite{Parisi1} they plot the average energy reached for fixed $J$ starting from $\caO(\sqrt{N})$ initial conditions. It is thus reasonable that \cite{contucci03} reports the lowest estimate of the energy at convergence. 
       \item Reluctant:  For the reluctant algorithm \cite{Parisi1} reports $e_{\rm reluctant} \approx -0.746$, while \cite[Figure 5]{contucci03} allows only to conjecture $e_{\rm reluctant} < -0.74$ (again, there is a difference in the averaging protocol, see bullet point above). In both \cite{Parisi1, contucci03} the authors consider $N \leq 1000$. 
        In this reluctant case, the literature that concludes that the algorithm does not asymptotically reach the ground state energy is less convincing as e.g. in \cite[Figure 5]{contucci03} it is clear that finite-size scalings are still important at $N \approx 300$. 
    \end{itemize}

    More elaborate versions of the greedy and reluctant algorithms have been considered, specifically in relation with the ``extremal optimisation'' heuristics \cite{BOETTCHER00}. In these variations to the algorithms the flipped spin is not the one with largest/smallest local field, but one chosen according to a probability distribution peaked on the one with largest/smallest local field. Moreover, for each instance of the disorder $J$ the algorithm is run for a given amount of trials (which may or may not scale with $N$ in different papers), and the minimum energy achieved is considered.
    \cite{BOETTCHER01} studies the greedy algorithm with these enhancements (probabilistic spin-flipping + restarts) and reports an energy density at convergence numerically which is compatible with $\egs$ when extrapolated to $N \to \infty$ (with simulations up to $N = 1023$).
    In the line of works by \cite{contucci05, contucci05-2}, a similarly enhanced version of the reluctant algorithms, and of an interpolation between reluctant and greedy, is studied, reporting convergence energies close but different from $\egs$, for system sizes up to $N \approx 300$. Note that many other complex heuristics approach exist in the literature (see e.g. \cite{bouchaud2003energy}).
    Avalanches in these kind of dynamics have been considered in \cite{baity2016criticality}.
        
    \subsection{Results on the single spin-flip dynamics}
    
    We performed extensive simulations for the barebones (no enhancements) random, greedy and reluctant algorithms, with the aim of better understanding these simple algorithms. In particular, the new numerical simulations allow us to clarify quantitatively the energy at convergence of all algorithms, going up to sizes $N = 40000$. 

    \begin{figure}[h]
        \centering
        \includegraphics[width=\textwidth]{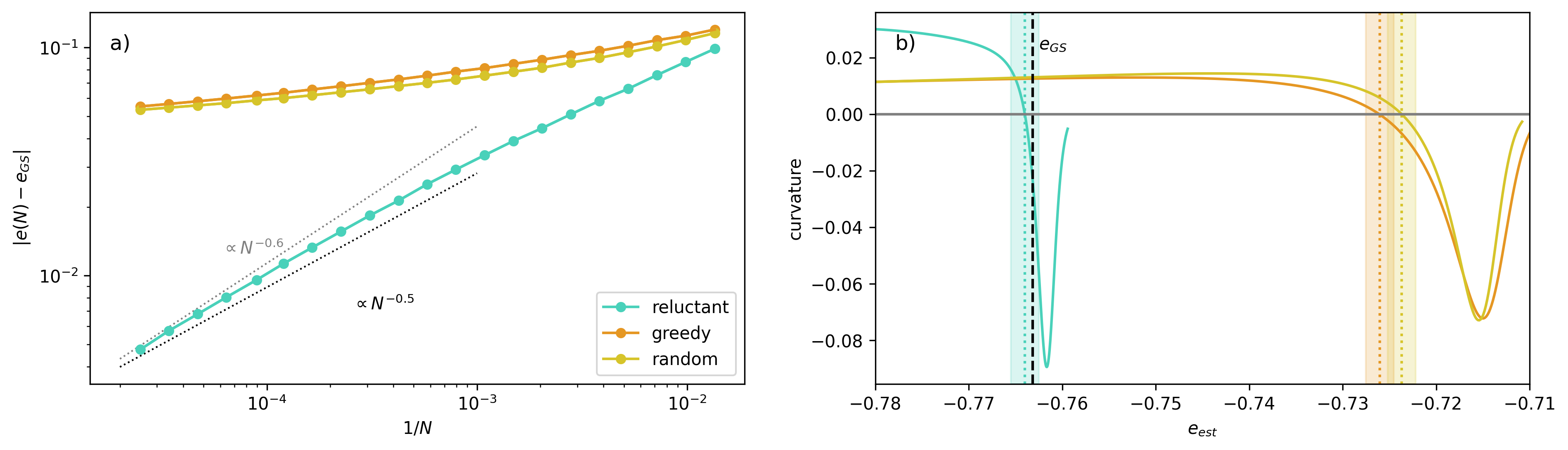}
        \caption{
        \textbf{Energy at convergence of the reluctant, greedy and random single spin-flip algorithms.} 
        (a) We plot the energy at convergence of the reluctant, greedy and random single spin-flip algorithms as a function of the system size $1/N$ in log-log scale.
        For the reluctant and greedy algorithms each datapoint is the average over 10000 samples, each with a newly sampled $J$ and newly sampled initial condition. For the reluctant algorithm we averaged over 3500 samples, but for the last four datapoints between $N=15581$ and $N=40000$ we only obtained between 1000 for the smaller and 64 for the larger size, due to the comparatively large complexity. 
        (b) To obtain a numerical estimate of the energy at convergence for all algorithms, we perform a quadratic fit of $\log | e_{\rm est} - e(N) |$ vs $\log N$, using the largest 8 sizes available. 
        We plot the estimated curvature (coefficient of the quadratic term of the fit) as a function of $e_{\rm est}$, and find the value of $e_{\rm est}$ such that the curvature is zero. This is then the estimate for the energy at convergence extrapolated to $N = \infty$. We shade error bars around the zero-curvature $e_{\rm est}$ of width $\pm 10^{-3}$ as we observed variability on the third digit when changing which points were used in the fitting procedure.
        We find 
        $e_{\rm est}^{\rm random} = -0.726 \pm 0.001$, 
        $e_{\rm est}^{\rm greedy} = -0.723 \pm 0.001$ and 
        $e_{\rm est}^{\rm reluctant} = -0.764 \pm 0.001$.
        }
        \label{fig:e_conv_vs_n}
    \end{figure}
    
    The main result concerns the energy at convergence of the reluctant algorithm. 
    We find that the energy at convergence $e(N)$ is compatible with $\egs$ when extrapolated $N \to \infty$.
    Figure \ref{fig:e_conv_vs_n}a, shows $\Delta e(N) = |e(N) - \egs|$ as a function of $1/N$ in log-log scale for all algorithms. 
    We observe that $\Delta e(N)$ of the greedy and random algorithms seem to plateau to a finite value as $N$ grows, while the $\Delta e(N)$ for the reluctant algorithm is compatible with $\lim _{N\to \infty} \Delta e(N) = 0$, and with a power-law finite-size scaling behaviour $\Delta e(N) \propto N^{-\alpha}$ with exponent $\alpha \approx 0.55 \pm 0.01$ (see the dotted lines in the plot).
    
    To obtain a numerical estimate of the energy at convergence of all algorithms at $N\to\infty$, we resort to the following fitting procedure.
    We consider the points $\log | e_{\rm est} - e(N) |$ vs $\log N$, where $e_{\rm est}$ is an estimate of the energy at convergence at $N = \infty$. If the finite size corrections are of power-law type and $e_{\rm est} = \lim_{N\to\infty} e(N)$, then $\log | e_{\rm est} - e(N) |$ vs $\log N$ should satisfy a linear relationship.
    Thus, we fit $\log | e_{\rm est} - e(N) |$ vs $\log N$ against a quadratic polynomial, and look for the value of $e_{\rm est}$ leading to a fit with zero curvature (i.e. zero coefficient of the second-order term).
    This provides an estimate of the energy at convergence $e(N)$ extrapolated to $N \to \infty$.
    Figure \ref{fig:e_conv_vs_n}b, shows the curvature (coefficient of a quadratic fit) of the datapoints $\log | e_{\rm est} - e(N) |$ vs $\log N$ as a function of $e_{\rm est}$, where we used only the 8 largest sizes to perform the fit. 
    We find 
    $e_{\rm est}^{\rm random} = -0.726 \pm 0.001$, 
    $e_{\rm est}^{\rm greedy} = -0.723 \pm 0.001$ and 
    $e_{\rm est}^{\rm reluctant} = -0.764 \pm 0.001$. 
    We estimate an error on the extrapolated energy at convergence on the third digit, as altering the fitting protocol (e.g. changing the amount of points used in the fit) leads to estimates that differ at most on the third digit.
    The estimate of the energy at convergence of the greedy algorithm is compatible with \cite{Horner07}. 
    The estimate of the energy reached by the reluctant algorithm is compatible with $\egs$ in disagreement with estimates from previous literature that were based on simulations for considerably smaller sizes.
    
    \begin{figure}[h]
        \centering
        \includegraphics[width=\textwidth]{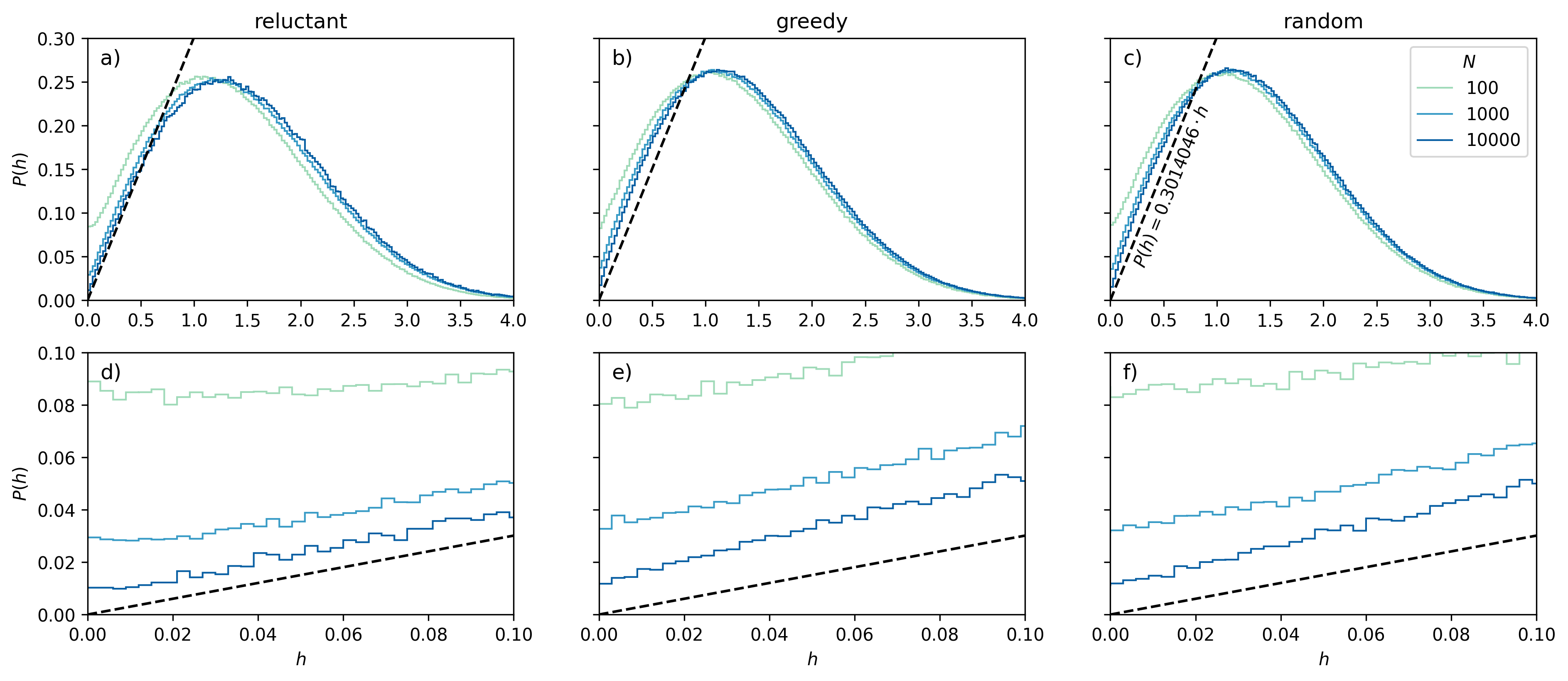}
        \caption{\textbf{Field distribution at convergence of the reluctant, greedy and random single spin-flip algorithms as a function of the system size $N$.} We compare the empirical distribution of the fields $h_i = N^{-1/2} \sum_{i=1}^N J_{ij} x_j$ at the convergence configuration $x$ of each algorithm. 
        The histograms are built buy sampling a new $J$ and a new initial condition several times, and binning them into 2000 equal-sized bins in $(-6,6)$. We plot only $h \in (0,6)$ (a,b,c) and $h \in (0, 0.1)$ (d,e,f) as the distribution is symmetric.
        For $N=100$ we generated 16000 samples, while for $N=1000$ and $N=10000$ we generated $1600$ samples. 
        The black dashed line represents the behaviour of the field distribution at the origin predicted for the SK ground state \cite{pankov}, i.e. $P(h) \sim \gamma h$ for $h \to 0^+$, with coefficient $\gamma \approx 0.3014$.
        }
        \label{fig:pseudogap}
    \end{figure}
    
    To corroborate the claim that the energy at convergence of the reluctant algorithm goes to the ground state, we studied the distribution of the local fields at convergence.
    Indeed, it is known that for the SK model the distribution of the local fields at the ground state is pseudo-gapped, i.e. its probability density function $P(h)$ is linear at the origin, $P(h) \sim \gamma h$ for $h \to 0^+$, with coefficient $\gamma = 0.301046\dots$ \cite{pankov} (see also \cite{Folena_2022} for a recent related discussion in a model of soft spins in a random anharmonic field).
    For completeness, the distribution is said to be not gapped if it is finite at $h=0$, pseudo-gapped if it is zero at the origin and linear in a neighbourhood of $h=0$, and gapped if it is zero in a finite interval around the origin.
   
    Figure \ref{fig:pseudogap} shows the empirical field distribution we computed from the numerical experiments for the random, greedy and reluctant dynamics for different sizes $N$.
    We find that the field distribution at convergence for all the three algorithms is compatible with the onset of a pseudo-gap, as the probability density of the fields at zero decreases as $N$ increases.
    We also see that only the reluctant dynamics seem to be compatible with the theoretical slope of the pseudo-gap, strengthening the claim that the reluctant algorithm is converging to the ground state. 
    The greedy and random algorithms feature a larger slope at the origin, i.e. a smaller pseudo-gap.

    In the field of disordered systems it has been predicted \cite{bouchaud1998out} that a slow simulated annealing algorithm with very slow cooling schedule also manages to reach configurations at energy compatible with $\egs$ in the SK model. 
    One could think that the reluctant algorithm, performing a slowest descent in energy, is behaving in some similar way to such a slow simulated annealing, and thus the two algorithms may be effective for related reasons. 
    While suggestive, this interpretation misses the crucial fact that simulated annealing can cross energy barriers due to thermal noise, while the reluctant algorithm follows trajectories along which the energy is monotonically decreasing. 
    This makes it even more surprising that the reluctant algorithm can reach the ground state energy.

    In Appendix \ref{app:numerical_e_est}, we provide additional information.
    We verify the convergence speed already reported in previous literature. The greedy and random algorithms converge in $\caO(N)$ spin flips, and the reluctant algorithm in $\caO(N^2)$ spin flips.
    We also compare these simple heuristics with a more sophisticated algorithm, incremental approximate message passing (IAMP) \cite{montanari,alaoui2020algorithmic}, which provably provides solutions with the ground state density in the thermodynamic limit with polynomial time convergence speed by leveraging the fully-replica-symmetry-broken solution of the SK model \cite{beyond}. We observe that for a given system size $N$, IAMP performs best among the algorithms we tested.
    In Appendix \ref{ap:3spin}, we present numerical results for the same dynamics in the 3-spin model that do not exclude the possibility that, also in that model, the reluctant algorithm achieves energies that are predicted to be the lowest achievable efficiently \cite{alaoui2020algorithmic,huang2022tight}.

    Analyzing the greedy and reluctant dynamics theoretically is challenging, even though they seem very simple dynamical systems.
    This is due to the fact that at each time step, the full set of $N$ local fields must be taken into consideration to find the largest/smallest unsatisfied spin to flip. This leads to a time-dependent choice of spin to flip that prevents us to use the common analytic tools used to characterise disordered dynamical systems.
    Providing an analytical characterisation of this class of dynamical systems is an important open problem.
    
    In the remainder of this work, we consider a related synchronous version of these update strategies, which are amenable to an analysis and are of independent interest. We will then draw some connections between single spin flip algorithms and synchronous algorithms in the conclusion.

    \section{Synchronous dynamics}\label{sec.sing}

    In order to obtain analytical insights into the dynamical processes aiming to reach local minima of the SK energy function, we shift our attention away from the single spin-flip dynamics, and instead consider a class of synchronous dynamics. For those at each time step a finite fraction of spins is flipped, in a way that allows for theoretical analysis.
    We consider the class of dynamical rules defined by
    \begin{equation}\label{eqdyn}
        x_i(t+1) = \sigma(x_i(t), h_i(t)) \mathfor i = 1, \dots, N \mathand t \geq 1 \, ,
    \end{equation}
    where $\sigma : \{ \pm 1 \} \times \bbR \to \{ \pm 1 \}$ is a generic function describing how the future state of a given spin $i$, namely $x_i(t+1)$, is determined by its current state $x_i(t)$, and by the collective state of the rest of the system through the local field $h_i(t)$. We denote discrete time by $t$, and set $t=1$ at initialisation.
    Even though the above dynamics is motivated as an analysable analogue for the single spin-flip algorithms, in their general form these dynamical rules are interesting by themselves. They describe the fully-connected analogue of outer-totalistic cellular automata \cite{Marr_2009,behrens2023dynamical} (usually considered on a grid or on sparse graphs).
    However, for this work we focus only on two rules that are inspired by the greedy and reluctant single spin-flip algorithms:
    \begin{itemize}
        \item Sync-greedy: the dynamical rule is
        \begin{equation}\label{eq-sync-greedy}
            \sigma_{\rm GR}(x, h; \kgr) = x \sign(x h + \kgr) \, .
        \end{equation} 
        Intuitively, for $\kgr > 0$, at each time step the dynamics flips all unsatisfied spins $x_i h_i < 0$ with large enough local field $|h_i| \geq \kgr$.
        For $\kgr < 0$, all unsatisfied spins are always flipped, as well as all satisfied spins whose local field is not strong enough. For example, the fixed points of the dynamics in this case are gapped configurations, where all spins are satisfied and the local fields are all stronger (in magnitude) than $\kgr$.
        \item Sync-reluctant: the dynamical rule is
        \begin{equation}
            \sigma_{\rm RL}(x, h; \krl) = x \sign(x h (x h + \krl)) \, .
        \end{equation}
        Intuitively, at each time step the dynamics flips all unsatisfied spins $x_i h_i < 0$ with weak enough local field $0 \leq |h_i| \leq \krl$. We consider here only the case $\krl > 0$, as for $\krl < 0$ one would flip only satisfied spins.
    \end{itemize} 
    
    Notice that the sync-greedy and sync-reluctant do not reduce to their single spin flip counterparts for any choice of the threshold $\kgr, \krl$, they are just different dynamical rules inspired by the single spin-flip algorithms. 
    In particular, they are not guaranteed anymore to converge to local minima as their single spin-flip counterparts, due to the energy being not necessarily monotone along the trajectories, and they can actually converge to attractors with periodicity $c > 1$.
    On the other hand, they both reduce to the dynamics $\sigma(x, h) = \sign(h)$ when either $\kgr = 0$ of $\krl = +\infty$. This is a classic dynamics considered for the SK model, a synchronous Glauber dynamics (we will refer to this as sync-Glauber in the following to stress its synchronous nature). 

    In the following, we study the convergence properties of the sync-greedy and sync-reluctant algorithms numerically and analytically, namely the energy density achieved at convergence and the properties of the dynamical attractors to which they converge.
    Before presenting results for the sync-greedy and sync-reluctant algorithms, we describe how one can study them analytically through Dynamical Mean Field Theory (DMFT), through the forward and backtracking version, the latter derived here for the first time for fully connected systems. 

    \paragraph{Related works.}

    The sync-Glauber dynamics, in particular for non-symmetric coupling, has been studied in detail in the literature, see for e.g. \cite{opper94, Henkel_1991, singh2003synchronous, hwang2019number}.
    As far as we know, the sync-greedy and sync-reluctant dynamics have not been studied previously in the literature. On the other hand, the fixed points of the sync-greedy dynamics, i.e. spin configurations such that $x_i h_i + \kgr > 0$, have been studied for negative value of $\kgr$ in the context of counting assortative and disassortative partitions on sparse graphs \cite{Behrens_assortative, minzer2023perfectly}, in the context of counting metastable states in the SK model \cite{braymoore, dandi2023maximallystable}, in the context of counting large margin fixed points of the sync-Glauber dynamics in dense associative memories \cite{Krauth_1987} and to study stability towards multiple spin flips in \cite{yan2015dynamics}. 
    
    A recent related work is \cite{garnierbrun2024unlearnable}, where the authors study with similar techniques a reinforcement learning dynamics in the SK model, possibly with non-symmetric coupling. As in our case, this is a synchronous dynamics aiming to minimise the energy.

    Regarding DMFT for the SK model, as far as we know it was first derived in \cite{opper94}. The backtracking DMFT we introduce below is inspired by the backtracking dynamical cavity method \cite{behrens2023backtracking} and by the cycle-counting procedure in \cite{hwang2019number}.

    \subsection{Dynamical Mean Field Theory}\label{sec.dmft}

    This section describes how DMFT allows one to predict for both the sync-greedy and sync-reluctant dynamics the following quantities of interest in the thermodynamic limit $N \to \infty$.
    \begin{itemize}
        \item Average energy density after $t$ time steps $e_{\rm fd}(t)$, starting from a random configuration. The average energy density is defined as
        \begin{equation}\label{defefd}
            e_{\rm fd}(t) = \lim_{N \to +\infty} \frac{1}{N} \EE_J \angavg{ H_J(x(t)) }_{\rm fd}  \mathfor t \geq 1\, ,
        \end{equation}
        where $x(t)$ is the spin configuration after $t-1$ times steps (i.e. $t=1$ is the initial condition), the angular brackets $\angavg{\cdot}_{\rm fd}$ denote an average over all spin trajectories starting from a random initial condition, and $\EE_J$ denotes the average over the disordered couplings.
        \item Annealed average energy inside a $(p,c)$-backtracking attractor $e^{p,c}_{\rm bk}(t)$. A $(p,c)$-backtracking attractor is a dynamical motif composed 
        by a
        transient of $p$ time steps which ends into a cycle of period equal to $c$ time steps. The energy $e^{p,c}_{\rm bk}(t)$ is defined analogously to $e_{\rm fd}(t)$, with the difference that the angular average $\angavg{\cdot}_{\rm bk}$ is an average over $(p,c)$-backtracking attractors, and $t$ counts time from the start of the transient of the backtracking attractor. 
        \begin{equation}\label{defefbwd}
            e^{p,c}_{\rm bk}(t) = \lim_{N \to +\infty} \frac{1}{N} \EE_J \angavg{ H_J(x(t)) }_{\rm bk} \mathfor 1 \leq t \leq p+c .
        \end{equation}
        We also consider $e^{p,c}_{\rm bk, min}$, the minimum average energy achieved in a $(p,c)$-backtracking attractor.
        When it is clear from the context, we drop the subscript on the angular brackets.
        \item Annealed entropy of $(p,c)$-backtracking attractors $s^{p,c}_{\rm bk}$. This entropy is a measure of how many initial conditions end up in a cycle of period $c$ after $p$ time steps. It is defined as
        \begin{equation}
            s^{p,c}_{\rm bk} = \lim_{N \to +\infty} \frac{1}{N} \log \EE_J |\text{Set of $(p,c)$-backtracking attractors} | \, ,
        \end{equation}
        and $| \cdot |$ denotes the cardinality of a set.
    \end{itemize}

    Before we study these observables in detail and compare to numerical simulations for both the sync-greedy and sync-reluctant dynamics, the remainder of this section discusses how to compute these quantities in the thermodynamics limit.
    
    \paragraph{The partition function.} To derive the DMFT predictions, we start by introducing a \textit{dynamical partition function}, defined as
    \begin{equation}\label{defZ}
        Z(\caT, J) = \sum_{\substack{\ux_1 \in \caT}}\sum_{\substack{\ux_2 \in \caT}}   \dots \sum_{\substack{\ux_N \in \caT}} \prod_{i=1}^N \prod_{t=1}^{T} 
        \delta\left[x_i(t+1) = \sigma\left( x_i(t), h_i(t) \right)\right]
    \end{equation}
    where $\ux_i = \{x_i(1), \dots, x_i(T+1)\}$ is the time evolution of spin $i$ for $T$ time steps, $\delta[a=b]$ is Kronecker's delta, and $\caT \subseteq \{\pm 1\}^{T+1}$ is a set of allowed single-spin trajectories.
    This partition function counts how many initial conditions $\{x_i(1)\}_{i=1}^N$ evolve under the deterministic dynamical rule $x_i(t+1) = \sigma\left( x_i(t), h_i(t) \right)$ and satisfy $\ux_i \in \caT$ for all spins $i=1, \dots, N$.
    As usual in statistical mechanics, we expect that $Z(\caT, J) = \exp(N s(\caT, J))$ where $s(\caT, J)$ is the associated entropy, and that the entropy will be determined by a maximisation principle over a set of order parameters characterising the system and its properties in the thermodynamic limit.
    Thus, our aim is to compute the entropy and the associated order parameters.\\
    We consider two cases:
    \begin{itemize}
        \item Forward DMFT, $\caT = \caT_{\rm fd} = \{\pm 1\}^{T+1}$. With this choice, the dynamical partition function describes an unconstrained system evolving for $T$ time steps from a random and uniformly drawn initial condition on $\{\pm 1 \}^N$. This is the classical version of DMFT \cite{opper94}.
        \item Backtracking DMFT, $\caT = \caT_{\rm bk}(p,c) = \{ (x(1), \dots, x(T+1)) \in \{\pm 1\}^{T+1} | x(p+1) = x(p+c+1) \}$ for $p + c = T$. 
        With this choice, the dynamical partition function \eqref{defZ} counts only initial conditions that after $p$ steps of the dynamics enter into a cycle of length $c$. 
        We call these dynamical motifs $(p,c)$-backtracking attractors, following \cite{behrens2023backtracking}.
    \end{itemize}
    To obtain the typical properties of the system averaged over the disorder, we consider in both cases the averaged partition function $\EE_J Z(\caT, J)$, and the corresponding annealed entropy
    \begin{equation}
        s_{\rm annealed}(\caT) = \lim_{N \to  \infty} N^{-1} \log \EE_J Z(\caT, J) \, .
    \end{equation}
    In order to describe the behaviour of typical instances of the problem, one should consider the quenched average $\EE_J \log Z(\caT, J)$. 
    However, for the forward DMFT annealed and quenched averages coincide as $Z(\caT_{\rm fd}, J) = 2^N$. For the backtracking DMFT we provide numerical experiments that
    hint towards the region of parameters for which annealed results predict the quenched properties of the system. Investigation of the actual quenched calculation is left for future work. 
    We also refer to \cite{hwang2019number}, where a sub-case of backtracking DMFT  ($p=0$, the case of cycle counting) was studied in the annealed approximation in a similar setting.
    
    \paragraph{Forward DMFT.} 
    In Appendix~\ref{app.dmft} we show that forward DMFT up to time $T+1$ is characterised by two sets of order parameters, $Q(t,s)$ and $V(s,t)$ for $1\leq t < s \leq T$.
    The order parameters are determined through a set of equations, namely
    \begin{equation}\label{eqspfd}
        \begin{split}
        Q(t,s) &= \angavg{x(s)x(t)}_{\ueta = 0}\, , 
        \quad
        V(s, t) =  \del_{\eta(t)} \angavg{x(s)}_\ueta \vert_{\ueta = 0} \mathfor 1\leq t < s \leq T\, ,
        \end{split}
    \end{equation}
    where the angular average is over trajectories of spins $\ux \in \{\pm 1\}^{T+1}$ and fields $\uh \in \bbR^T$  distributed as
    \begin{equation}
        p^{\rm fd}_\ueta(\ux, \uh) = \frac{1}{2} \caN\left( \uh ; \sum_{s<t} V(t,s)x(s), \delta_{t,s} + Q(t,s) \right) 
        \prod_{t=1}^{T} 
        \delta\left[x(t+1) = \sigma\left( x(t), h(t) +  \eta(t) \right)\right]
        \, ,
    \end{equation}
    where $\caN(x; \mu, \Sigma)$ is the density of a multivariate Gaussian with mean $\mu$ and covariance $\Sigma$, and $\ueta$ is a fixed perturbation to the fields $\uh$ (notice that $p^{\rm fd}_\ueta(\ux, \uh)$ is normalised only for $\ueta = 0$). 
    % In the equations above, $\del_{h(t)}$ inside an angular average denotes the derivative with respect to $h(t)$ of the product of delta functions in the distribution $p(\ux, \uh)$.
    At the solution of \eqref{eqspfd}, the observables are given by
    \begin{equation}
        \begin{split}
            e_{\rm fd}(t) = - \sum_{s<t} V(t,s) Q(s,t)
            \mathand
            s_{\rm annealed} = s_{\rm quenched} = \log 2 \, .
        \end{split}
    \end{equation}
    The distribution $p^{\rm fd}(\ux, \uh)$ can be regarded as a measure defining a stochastic process $(\ux, \uh)$, modelling a mean-field approximation to the dynamics of the spins and local fields in the real system. 
    In this light, $Q$ is the time-to-time spin-spin correlation function, and $V$ is the response function quantifying how a perturbation over the local field at time $t$ affects the spin dynamics at later times.

    In Appendix~\ref{app.fd} we show that the stochastic process is causal, leading to great simplifications in the numerical evaluation of the order parameters.
    By causal we mean that at a given time $t$, the spins $x(t)$ and fields $h(t)$ are statistically independent from  $x(t'), h(t')$ for all future times $t' > t$ under the measure of the stochastic process. In other words, the stochastic process can be sampled by sampling, one time step after the other, $x(1)$, $h(1)$, $x(2)$, $h(2)$, etc \dots
    This in turn implies that the equations \eqref{eqspfd} are not self-consistent: for a given time $t$, all the order parameters $\{Q(t,s), V(s,t)\}_{s < t}$ depend only on order parameters $\{Q(t',s), V(s,t')\}_{s < t'}$ at previous times $t' < t$.
    Thus, the order parameters can be computed time-by-time recursively. The only practical limitation is the size of the order parameters scaling as $\caO(T^2)$ in the number of time steps $T$. Also, \eqref{eqspfd} can be efficiently evaluated by Monte Carlo integration as the stochastic process is easy to sample.
    These observations allow us to compute the order parameters up to times of the order $T \approx 400$, and with further optimizations this could even be improved (see Appendix~\ref{app.numericsdmft}).

    Before moving to the backtracking DMFT, we recall some results for the forward DMFT of the sync-Glauber dynamics. 
    In \cite{opper94, Parisi2} it is shown that, for the forward DMFT, $V(t,s) = 0$ if $t,s$ have same parity, and $Q(t,s) = 0$ for $t,s$ with opposite parity. This readily implies that the average energy after $t$ steps from random initial condition is zero for any finite $t$. Moreover, the dynamics always converges to 2-cycles.

    \paragraph{Backtracking DMFT.} 
    The backtracking DMFT is more involved. In Appendix~\ref{app.dmft} we show that backtracking DMFT up to $T = p+c$ time steps is characterised by three sets of order parameters $Q, V$ and $R$, determined through the set of self-consistent equations
    \begin{equation}\label{eqspgeneral}
        \begin{split}
        Q(t,s) &= \angavg{x(t)x(s)}_{\ueta = 0} \mathfor 1\leq t<s \leq T \, , 
        \\
        V(t,s) &= \del_{\eta(t)} \angavg{x(s)}_\ueta \vert_{\ueta = 0} \mathfor 1\leq t,s \leq T\, ,
        \\ 
        R(t,s) &= \del^2_{\eta(t),\eta(s)} \angavg{1}_\ueta \vert_{\ueta = 0} \mathfor 1\leq t<s \leq T \, ,
        \end{split}
    \end{equation}
    where the angular average is over the distribution of spins and fields
    \begin{equation}
        p^{\rm bk}_\ueta(\ux, \uh) = \frac{e^{\sum_{t<s}R(t,s)x(t)x(s)} }{\caZ} \caN\left( \uh ; \sum_s V(t,s)x(s), \delta_{t,s} + Q(t,s) \right) 
        \prod_{t=1}^{T} 
        \delta\left[x(t+1) = \sigma\left( x(t), h(t) + \eta(t) \right)\right]
        \, ,
    \end{equation}
    with $\ux \in \caT_{\rm bk}(p,c)$, $\uh \in \bbR^T$ and $\caZ$ is a non-trivial normalisation factor (again, notice that $p^{\rm bk}_\ueta(\ux, \uh)$ is normalised only for $\ueta = 0$, so that the equation for $R$ is non-trivial).
    The observables are given by
    \begin{equation}
        \begin{split}
            e^{p,c}_{\rm bk}(t) = - \sum_s V(t,s) (\delta_{t,s} + Q(t,s))
            \mathand
            s^{p,c}_{\rm bk, annealed} 
            &= \log\caZ - \sum_{t<s} Q(t,s) R(t,s) + \frac{1}{2} \sum_{s,t} V(t,s) V(s,t) \, .
        \end{split}
    \end{equation}    
    As before, $p^{\rm bk}(\ux, \uh)$ defines a stochastic process, with the same interpretation as in forward DMFT. This time, the conditioning over the end of the dynamics being a cycle of length $c$ breaks causality, introducing in the stochastic process a statistical dependence between variables at all times which prevents straightforward sampling as we could do in forward DMFT. 
    Moreover, \eqref{eqspgeneral} are self-consistent, meaning that they require numerical solution, and not just recursive numerical evaluation as in the case of forward DMFT. This makes it computationally difficult to obtain the values of the order parameters for total times $T=p+c > 9$, at least using our numerical implementation (see Appendix~\ref{app.numericsdmft} for more details).

    We remark that, for example, the entropy of $(p, c=2)$-backtracking attractors automatically counts also $(p, c=1)$-backtracking attractors, as cycles of length one are trivially also cycles of length two. 
    We show in Appendix~\ref{app.degeneracy} that \eqref{eqspgeneral} for $c=2$ exhibits in general two distinct solutions, one inherited from the $c=1$ solution, and another non-trivial one describing proper $c=2$ backtracking attractors. The correct solution to compute observables is the one leading to larger entropy.
    This kind of multiplicity of solutions exists also for larger values of $c$, and was already observed in \cite{hwang2019number}, where the authors discuss in detail the problem of isolating the count of proper $c$-cycles from their trivial counterparts composed by smaller cycles.

    We also point out that another type of trivial $c=2$ backtracking attractor exists, ending in what we call a \textit{fully-rattling} 2-cycle, where the ending 2-cycle satisfies $x(p) = - x(p+1) = x(p+2)$. These 2-cycles can still be analyzed in the backtracking DMFT framework with minimal modification, namely by considering the $(p, c=1)$ entropy with anti-periodic condition $x(p) = - x(p+1)$.

    \paragraph{Field distribution.}
    We highlight that the stochastic process $p(\ux, \uh)$ gives us also theoretical access to the distribution of the fields at a given time, i.e. $\text{Prob}(h_i(t) = y)$, and to the distribution of $\text{Prob}(x_i(t) h_i(t) = y)$, simply by marginalising over all other spin and field variables.
    
    \paragraph{}
    In the following sections we compare numerical estimates of the observables for the sync-greedy and sync-reluctant dynamics with the analytical results of the DMFT as outlined previously.

        We remark that the DMFT equations we derive in Appendix~\ref{app.dmft} generalise without any modification to any dynamical rule as in \eqref{eqdyn}, and to non-symmetric couplings $J$. Generalisation to stochastic dynamical rules is also possible with minimal modifications. Also, as usual for the SK model, all the presented results generalise without alteration to any distribution of the couplings $J$ with finite first and second moments matching those for the Gaussian case we consider here.

    \subsection{The sync-greedy algorithm}

    The sync-greedy algorithm has the dynamical rule $\sigma_{\rm GR}(x, h; \kgr) = x \sign(x h + \kgr)$: all spins $i$ such that $x_i h_i < - \kgr$ are flipped at each timestep.
    In the following, we aim to understand to which attractors the dynamics converges, what energy these attractors have, and how long it takes to converge to them.
    
    \paragraph{Properties of the dynamics.}
    The sync-greedy dynamics has simple attractors: for any set of couplings $J$, it can only converge to cycles of length $c=1,2$ \cite{goles2cycles}.
    This makes it particularly easy to numerically detect convergence to an attractor, and allows us to study backtracking DMFT only for $c=1,2$.
    
    \begin{figure}
        \centering
        \includegraphics[width=0.95\textwidth]{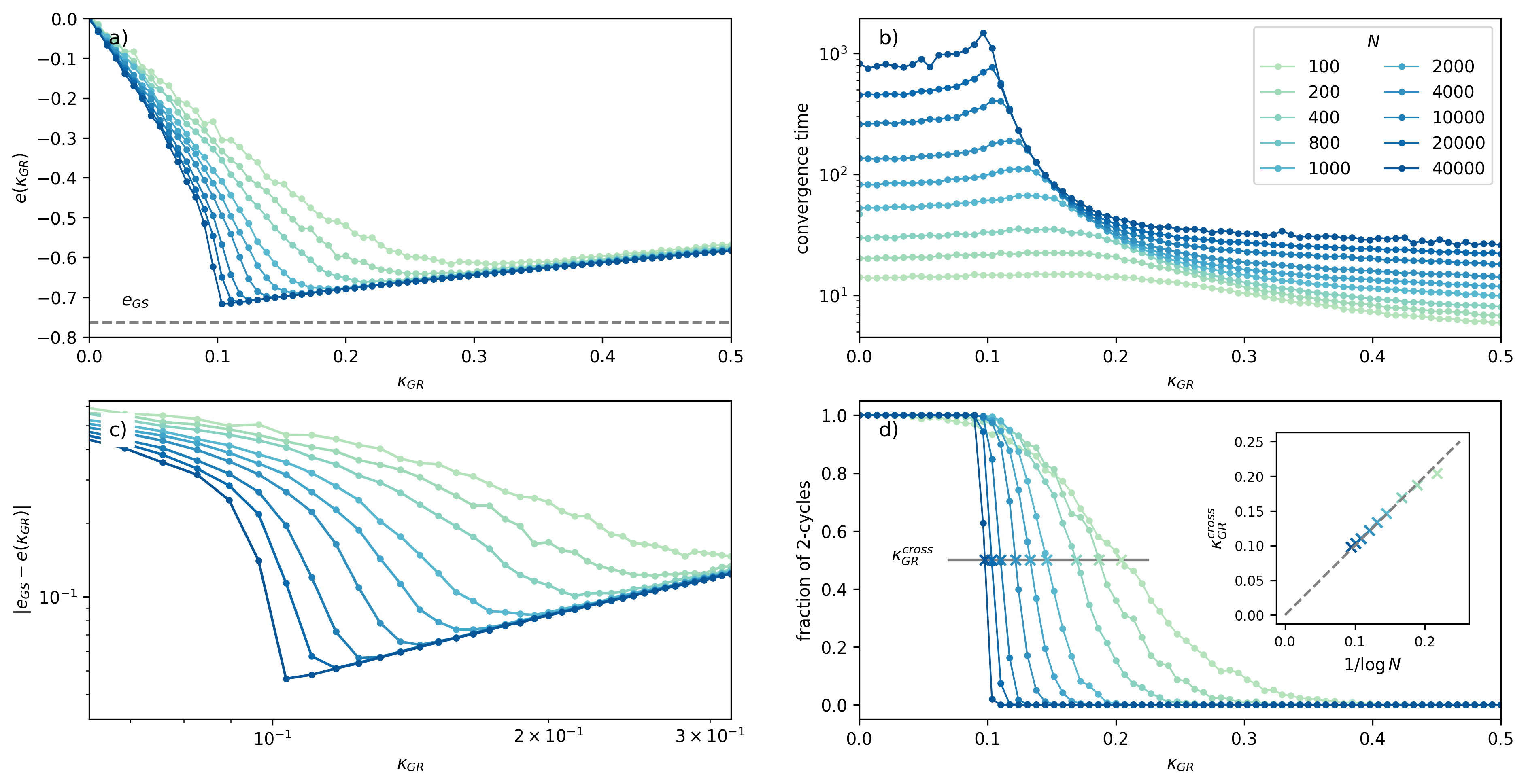}
        \caption{
        \textbf{Sync-greedy dynamics - Numerical experiments.}
        \textit{(a)} Energy at convergence, \textit{(b)} convergence time and \textit{(d)} average final cycle length for increasing values of $N$ and $\kgr \in [0,0.5]$. 
        Each datapoint is the average over $800$ runs. 
        At finite size $N$ we can clearly recognise two regimes: for small $\kgr$ the dynamics is dominated by 2-cycles with not so low energy, while for large $\kgr$ the dynamics is dominated by 1-cycles with much lower energy.
        \textit{(c)} Energy at convergence minus $\egs$ as a function of $\kgr$ in log-log scale (same data as in panel a).
        \textit{(d, inset)} We approximate the transition between $1$-cycles and $2$-cycles, i.e. $\kcgr(N)$, as the
        value of $\kgr$ for which the dynamics converges with probability $1/2$ to fixed points of 2-cycles (and the other half to 1-cycles). We plot $\kcgr(N)$ vs $1/\log N$ in the inset, and see that it is compatible with $\lim_{N \to \infty} \kcgr(N) = 0$. The dashed line is the diagonal $\kcgr(N) = 1/\log N$.
        }
        \label{fig:sync-eo-empirics}
    \end{figure}

    While we define the sync-greedy dynamics for an arbitrary real valued $\kgr$ (positive or negative) we notice a symmetry thanks to which  
    we can study the dynamics only for $\kgr \geq 0$ and obtain the results for $\kgr < 0$ from that.
    Consider a trajectory $x_i(t)$ satisfying the dynamical update rule \eqref{eq-sync-greedy} with disorder $J$ and threshold $\kappa$. 
    Then, one can see that the trajectory $y_i(t) = (-1)^t x_i(t)$ satisfies the same update rule with disorder $-J$ and threshold $-\kappa$. Thus, on average over the disorder (which is symmetrically distributed), all properties of the $\kgr < 0$ dynamics can be mapped to properties of the $\kgr > 0$ dynamics. 
    Under this mapping, we have the following equivalences: fixed points are mapped into fully-rattling $c=2$ cycles and \textit{viceversa}, non-trivial $c=2$ cycles are mapped into non-trivial $c=2$ cycles, and the energy is mapped into minus itself. Thus, it suffices to study the system at $\kgr \geq 0$ to obtain the picture for $\kgr \in {\mathbb R}$.

    We also remark that for $\kgr \to -\infty$ all configurations are fully-rattling $c=2$ cycles, for $\kgr = 0$ the dynamics converges to non-trivial $c=2$ cycles (as it reduces to the sync-Glauber dynamics), and for $\kgr \to +\infty$ all configurations are fixed points. 
 
    \paragraph{Numerical experiments.}
    Fig.~\ref{fig:sync-eo-empirics} summarizes our numerical experiments, for sizes up to $N = 40000$, as a function of the threshold parameter $\kgr$. We focus on $\kgr\ge 0$. The experiments are performed by simply running the dynamics from random initial conditions up to convergence, and averaging over $800$ samples (each time resampling both the disorder $J$ and the initial condition). We measured the energy at convergence (Fig.~\ref{fig:sync-eo-empirics}a), the convergence time (Fig.~\ref{fig:sync-eo-empirics}b), and the length of the cycles to which the dynamics converge (Fig.~\ref{fig:sync-eo-empirics}d).
    
    At finite size $N$ we observe two regimes delimited by a size-dependent threshold $\kcgr(N)$ (the threshold is plotted in Fig.~\ref{fig:sync-eo-empirics}d, inset). 
    For $\kgr < \kcgr(N)$, the dynamics converges more often to 2-cycles.
    The 2-cycles have an energy sensibly larger than $\egs$.
    For $\kgr > \kcgr(N)$, the dynamics converges more often to 1-cycles, which have an energy only slightly larger than $\egs$. 
    Around the threshold $\kgr \approx \kcgr(N)$, convergence time is largest, and we numerically observe cycles of both length 1 and 2. The average energy reached at convergence is lowest around $\kgr \approx \kcgr(N)$, and relatively close to $\egs$.
    Thus, as $\kgr$ increases, we observe a cross-over between a dynamics dominated by 2-cycles with higher energy, and 1-cycles with much lower energy, closer to the ground state energy.
    The cross-over becomes sharper as the system size $N$ increases.
    The behaviour at $\kgr < \kcgr(N)$ is compatible with the behaviour at $\kgr = 0$, where the dynamics reduces to the sync-Glauber dominated at convergence by 2-cycles of zero energy (see Section~\ref{sec.dmft}).
    
    It is natural to wonder whether $\lim_{N \to \infty} \kcgr(N)$ is zero, or remains finite. 
    The inset in Fig.~\ref{fig:sync-eo-empirics}d shows $\kcgr(N)$ as a function of $1/\log(N)$. 
    We estimated $\kcgr(N)$ as the value of $\kgr$ for which on average the dynamics converges half of the times to 1-cycles, and half of the times to 2-cycles.
    We see that, in the range of sizes that we could access numerically, $\kcgr(N)$ is compatible with a $1/\log(N)$ decay to zero, leading us to conjecture that $\lim_{N \to \infty} \kcgr(N) = 0$.
    This would imply a discontinuous phase transition at $\kgr = 0$, from the $\kgr < 0$ phase dominated by fully-rattling 2-cycles, to the $\kgr > 0$ phase dominated by convergence to 1-cycles of energy close to $\egs$.
    
    In Fig.~\ref{fig:sync-eo-empirics}c we plot the difference between the energy at convergence and the ground state energy $e - \egs$ as a function of $\kgr$ for different system sizes, and show that the energy difference for $\kgr > \kcgr(N)$
    is compatible with going to zero as $N \to \infty$ (it has a linear behaviour on a log-log scale). This together with the conjecture $\lim_{N \to \infty} \kcgr(N) = 0$ would imply that the sync-greedy dynamics could also go to the ground state energy when $N\to +\infty$ and $\kcgr \to 0^+$ (in this order).
    
\begin{figure}
        \centering
        \includegraphics[width=\textwidth]{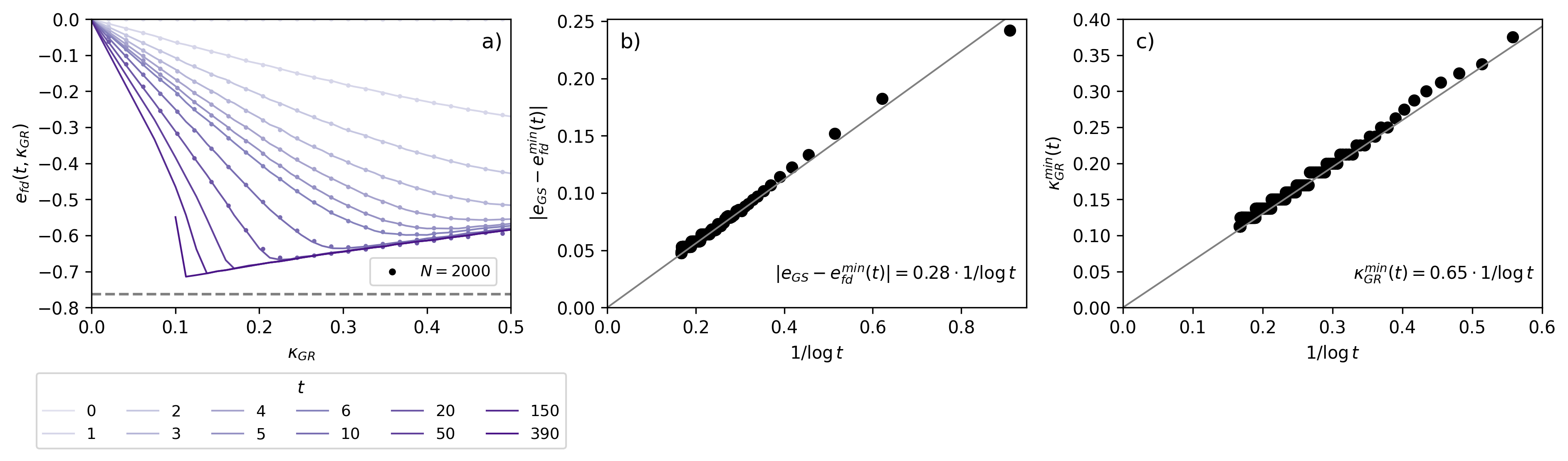}
        \caption{
        \textbf{Sync-greedy dynamics -- Forward DMFT.} 
        \textit{(a)} 
        Average energy after $t$ time steps from random initialisation, obtained using forward DMFT for $t \lessapprox 400$. The markers represent the empirical energy on instances of size $N=2000$ after $t$ time steps, where each data point is the average over $3000$ runs (at each run, we resample both $J$ and the initial condition). The solid lines are the DMFT prediction.
        \textit{(b)} 
        Difference between $\egs$ and the minimum energy reached after $t$ forward time steps as predicted by forward DMFT, $e^{\rm min}_{\rm fd}(t)$, vs $1 / \log t$. 
        The grey reference line is defined in the plot.
        \textit{(c)} 
        Value of the threshold $\kgr$ at which the minimum energy is reached after $t$ forward time steps as predicted by forward DMFT, $\kmingr(t)$, vs $1 / \log t$. 
        The grey reference line is defined in the plot.
        }
        \label{fig:sync-eo-dmft-fd}
    \end{figure}

    \paragraph{Forward DMFT.}
    To better understand the numerical observations, we analyse the dynamics using DMFT. Fig.~\ref{fig:sync-eo-dmft-fd} summarises the results of our analysis for the forward DMFT.

    In Fig.~\ref{fig:sync-eo-dmft-fd}a we show the average energy after $t$ steps from a random initialisation $e_{\rm fd}(t)$ (solid lines are the $N \to \infty$ DMFT predictions, and dots are numerical experiments at $N=2000$). 
    We see that for any fixed $\kgr \gtrapprox 0.1$ the energies reached by the DMFT converged very fast to their large time limits (indeed e.g. the energy at step 390 is indistinguishable from the one at time 150 for all $\kgr \gtrapprox 0.13$),
    
    Via the forward DMFT, we can also access the correlation between the spin configurations a two different time steps
    \begin{equation}
        \lim_{N \to \infty} \frac{1}{N} \sum_i \EE_J \angavg{ x_i(t) x_i(s) }_J = Q(t,s)
    \end{equation}
    where $Q(t,s)$ is one of the order parameters introduced in Section \ref{sec.dmft}, the time-to-time spin-spin correlation function. 
    By looking at $Q(T-1, T)$, where $T$ is an estimate of the convergence time, we can detect the type of attractor to which the dynamics converges: $Q(T-1, T) \approx 1$ (within numerical precision) implies convergence to cycles of length 1, and $Q(T-1, T) < 1$ implies convergence to cycles of length 2. Up to the precision of the numerical solution of the forward DMFT equations, we observed that $Q(T-1, T) \approx 1$ for all $\kgr \gtrapprox 0.1$, where forward DMFT reaches convergence in $t \lessapprox 400$ time steps, indicating that for all such values of $\kgr$ the dynamics converges to fixed points (see Appendix Fig.~\ref{fig:app:correlationfd}). We used the largest $T = 390$ available for each $\kgr$.
    
    Let us define $\kmingr(t)$ as the value of $\kgr$ at which forward DMFT after $t$ steps reaches the minimum energy, and $e^{\rm min}_{\rm fd}(t)$ the corresponding value of the minimum energy. We plot $\kmingr(t)$ and $|e^{\rm min}_{\rm fd}(t) - \egs|$ vs $1 / \log t$ in Fig.~\ref{fig:sync-eo-dmft-fd}b and  Fig.~\ref{fig:sync-eo-dmft-fd}c, and see that as $t$ grows $\kmingr(t)$ is compatible with $\lim_{t \to +\infty} \kmingr(t) = 0$, which supports the conjecture that the dynamical phase transition happens at $\kgr \to 0^+$.
    Moreover, we see again that the minimum energy achieved is compatible with $\lim_{t \to +\infty} e^{\rm min}_{\rm fd}(t) = \egs$.
    
    Thus, is seems that the phase $\kgr > \kmingr(t)$ is akin to the phase $\kgr > \kcgr(N)$, and for both thresholds we conjectured that they go to 0 respectively for $t \to \infty$ or for $N\to \infty$.

\begin{figure}
        \centering
        \includegraphics[width=\textwidth]{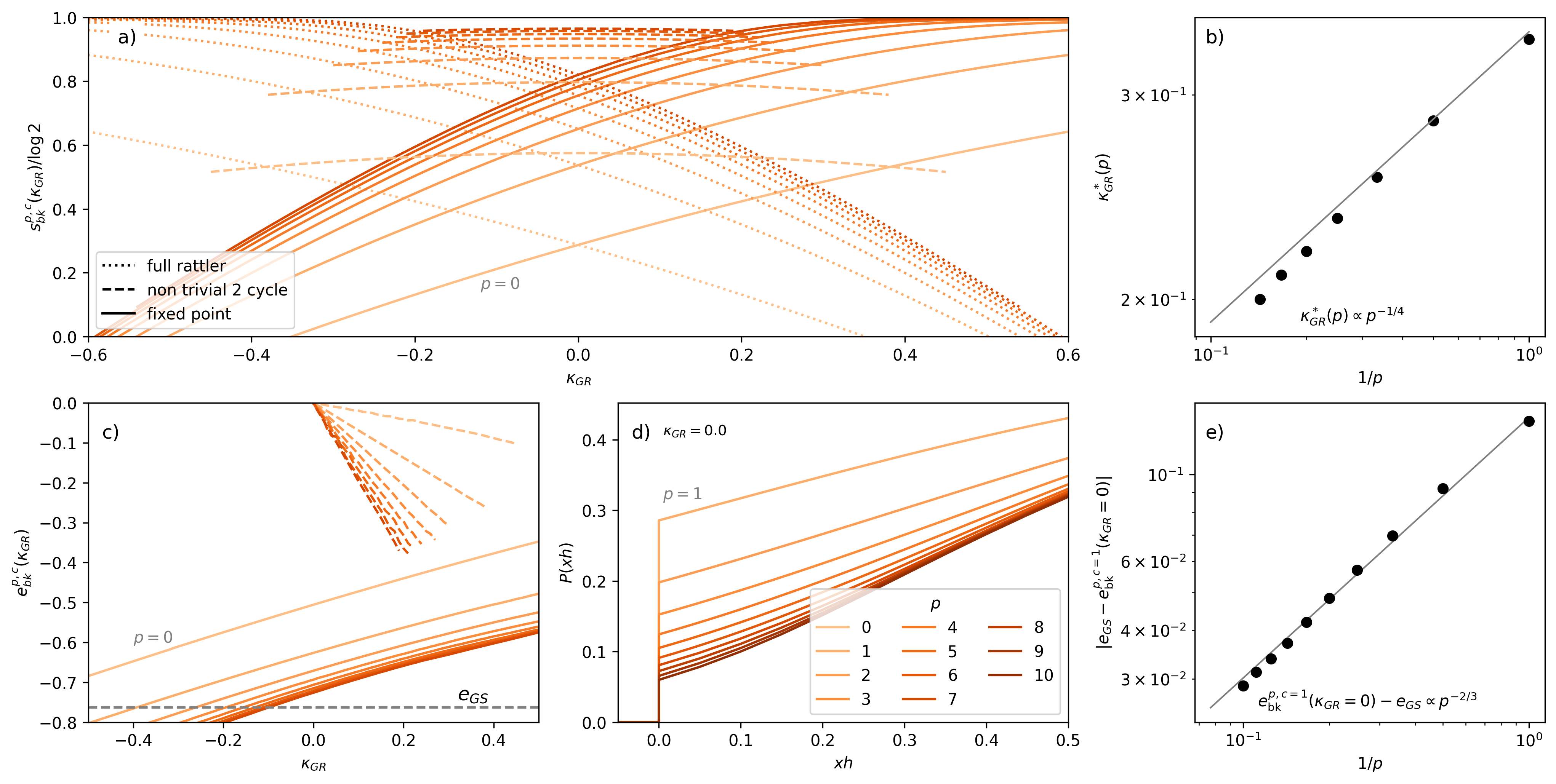}
        \caption{
        \textbf{Sync-greedy dynamics -- Backtracking DMFT.} 
        \textit{(a)} 
        Normalized entropy $s^{p,c}_{\rm bk} / \log 2$ as a function of $\kgr$, obtained using backtracking DMFT for cycles of length $c=1$ (solid), length $c=2$ (dashed), and fully-rattling length $c=2$ (dotted), for different transient lengths $0 \leq p \leq 7$. 
        \textit{(b)} 
        Critical threshold $\ksgr(p)$, separating the region where $c=1$ attractors dominate from that in which $c=2$ attractors dominate, as a function of $1/p$.
        The grey line is a reference power-law, as indicated in the plot.
        \textit{(c)} Average energy in a $(p,c)$-backtracking attractor $e^{p,c}_{\rm bk}$ as a function of $\kgr$, obtained using backtracking DMFT for cycles of length $c=1$ (solid) and length $c=2$ (dashed) for different transient lengths $0 \leq p \leq 7$. Fully-rattling 2-cycles have positive energy for all $\kgr$, and similarly $c=2$ non-trivial cycles for $\kgr <0$. Thus, we do not plot them here.
        \textit{(d)} 
        Probability density function at the fixed point $P(x(t) h(t))$ in a $(p,c=1)$-backtracking attractor for $\kgr = 0$ for different values of $p$. We highlight the region around the origin, showing the onset of a pseudo-gap as $p$ increases.
        \textit{(e)}
        Average energy in a $(p,c=1)$-backtracking attractor for $\kgr = 0$ as a function of $1/p$. The grey line is a reference power-law, as indicated in the plot. 
        }
        \label{fig:sync-eo-dmft-bk}
    \end{figure}

    \paragraph{Backtracking DMFT.}
    In Fig.~\ref{fig:sync-eo-dmft-bk} we show the results from backtracking DMFT, focusing on $(p,c)$-backtracking attractors with $c=1,2$ and fully-rattling $c=2$ attractors for $0 \leq p \leq 7$. 
    In Fig.~\ref{fig:sync-eo-dmft-bk}a we plot the entropy (normalised by its maximum value $\log(2)$) of $(p,c)$-backtracking attractors $s^{p,c}_{\rm bk}(\kgr)$. 
    For fixed $p$, as $\kgr$ grows, we observe three phases, delimited by $\kgr = \pm \ksgr(p)$: for $\kgr < - \ksgr(p)$, fully rattling $c=2$ attractors have the largest entropy, and dominate the behaviour of $(p,c)$-backtracking attractors.
    For $- \ksgr(p) < \kgr < \ksgr(p)$, non-trivial 2-cycles dominate, and for $\kgr > \ksgr(p)$ fixed points dominate.    
    This is compatible with the fact that the dynamics at $\kgr = 0$ converges to 2-cycles (see Section \ref{sec.dmft}), as well as with the behaviour at $\kgr = \pm \infty$. The threshold $\ksgr(p) > 0$ marks a discontinuous phase transition for any fixed $p$, but it seems to decrease as we increase $p$ (Fig.~\ref{fig:sync-eo-dmft-bk}b). 
    In other words, the phase in which non-trivial 2-cycles dominate shrinks.
    
    Similarly as for $\kcgr(N)$ and $\kmingr(t)$, we conjecture that $\lim_{p \to \infty} \ksgr(p) = 0$, fitting once again into our picture that a dynamical phase transition happens at $\kgr = 0$.
    We also notice that as $p$ increases the normalised entropy gets closer and closer to 1, meaning that more and more initial conditions do converge to $c=1,2$ cycles after just $p=7$ steps (even though this is still a subleading fraction of the total amount of initial conditions). This gives confidence that the results from backtracking DMFT are actually describing the typical convergence properties of the randomly initialized dynamics.

    In Fig.~\ref{fig:sync-eo-dmft-bk}c we plot the average energy found in $(p,c)$-backtracking attractors (for 2-cycles, averaged between the two configurations in the attractor). We see that the 2-cycles have energies far from $\egs$, while 1-cycles have much lower energies, getting closer and closer to $\egs$ as $\kgr$ decreases.
    It is natural to wonder whether at $\kgr = 0^+$ the subdominant 1-cycles reach an energy which, extrapolated to $p \to +\infty$, is compatible with $\egs$. A positive answer, in conjunction with our conjecture of a dynamical phase transition at $\kgr = 0$, would mean that for vanishingly small $\kgr$ the dynamics converges to fixed points at $\egs$ (even though with large convergence times), as we already started arguing above through numerical experiments and forward DMFT.
    In Fig.~\ref{fig:sync-eo-dmft-bk}e, we plot $e_{\rm bk}^{p,c=1}(\kgr = 0)$ as a function of $p$, and see that it is compatible with a power-law convergence to $\egs$ for $p \to \infty$. 
    In Fig.~\ref{fig:sync-eo-dmft-bk}d, we plot the distribution $P(x(t)h(t))$ in $(p,c=1)$-backtracking attractors for different values of $p$, and observe that the value of the p.d.f. at $xh =0$ decreases, compatibly with the onset of a pseudo-gap characteristic of the ground state.

    We remark that the extrapolation of $e_{\rm bk}^{p \to \infty,c=1}(\kgr = 0) = \egs$ and $\ksgr(p \to \infty) = 0$ has to be considered only as a rather weak evidence for our conjecture. Indeed, we are extrapolating from $0 \leq p \leq 10$ to $p=\infty$, so it is plausible that an asymptotic regime may have not been reached yet. Moreover, backtracking DMFT in the annealed approximation provides a lower bound to the energy computed using the quenched average, so that the true $e_{\rm bk}^{p,c=1}(\kgr = 0)$ may be larger \textit{a priori}. We consider this latter aspect more carefully in the next paragraph. 

    \paragraph{Comparison between numerical experiments and the backtracking DMFT, and validity of the annealed approximation.}

    All results shown above from DMFT are obtained in the annealed approximation. This is justified for the forward DMFT, as in that case quenched and annealed averages lead to the same entropy, but the results may be just an approximation (upper bound on the quenched entropy) in the backtracking DMFT case.
    Here we assess the validity of the results from backtracking DMFT by comparing with numerical experiments for the properties of $(p,c)$-backtracking attractors.

Here we will aim to sample $(p,c)$-backtracking attractors, and compute their properties. For a given disorder $J$ a random initial configuration will only lead to an attractor of length $c$ exponentially rarely in the system size. We can still do a numerical experiment where for small sizes we run exponentially many trials to probe the properties of $(p,c)$-backtracking attractors. Atypical disorder configurations $J$ may have many more $(p,c)$-backtracking attractors than typical graphs. We thus can compute the two following averages:     
    \begin{itemize}
        \item Annealed sampling: we sample a new instance of the disorder $J$ and a new initial condition, run the dynamics for $p$ steps and check if it reached a cycle of length $c$. If it did, the sample is kept and used to compute averages, otherwise it is discarded. 
        \item Quenched sampling: we sample a new instance of the disorder $J$. Then, we sample a new initial condition, run the dynamics for $p$ steps and check if it reached a cycle of length $c$. This is repeated with the same disorder $J$ until one initial condition reaches a cycle of length $c$. Only then the sample is kept, and $J$ is regenerated. 
    \end{itemize}
    Both procedures are repeated until a given number of $(p,c)$-backtracking attractors has been found to then compute their average properties.

 \begin{figure}
        \centering
        \includegraphics[width=\textwidth]{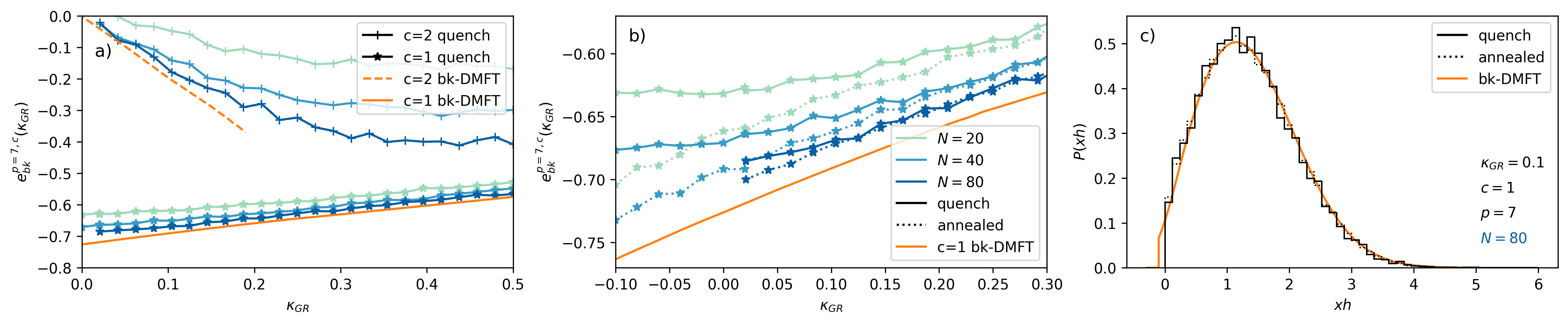}
        \caption{
        \textbf{Sync-greedy dynamics -- Numerical experiments vs annealed backtracking DMFT.} 
        \textit{(a)} 
        Average energy in a $(p=7,c)$-backtracking attractor $e^{p,c}_{\rm bk}$ as a function of $\kgr$, for $c=1,2$. Orange lines are the theoretical prediction of backtracking DMFT in the annealed approximation. 
        Colored lines/dots are numerical experiments for sizes $N=20,40,80$. Each point is the quenched average over 1000 realisations of the disorder. For each realisation of the disorder, we sampled initial conditions until one was found that converged to a cycle of length $c$ after $p=7$ steps. We see that the numerical experiments get closer and closer to the prediction of the backtracking DMFT as $N$ increases.
        \textit{(b)} 
        Same plot as in the left panel, but focusing only on $c=1$. This time, two sets of numerical experiments were performed. One analogous as in the left panel, corresponding to a quenched empirical average. The other, corresponding to an annealed empirical average, has been generated by averaging over 1000 samples for $N=20,40$, each time regenerating both the disorder $J$ and the initial condition. We see that annealed and quenched empirical averages converge towards the theoretical prediction of the annealed backtracking DMFT as $N$ increases. 
        We also observe that for smaller values of $\kgr$ the discrepancy is getting larger.  For $N=80$ with extensive sampling we find 200 instances of the disorder. For $\kgr < 0.02$ we do not show any data, as it contained too few samples.
        \textit{(c)}
        Comparison between the distribution of $x(t) h(t)$ from annealed experiments, quenched experiments and annealed backtracking DMFT for $\kgr = 0.1$, $N=80$, $c=1$ and $p=7$. We observe a nice match.
        }
        \label{fig:annealed-quenched}
    \end{figure}

    In Fig.~\ref{fig:annealed-quenched} we compare the annealed backtracking DMFT predictions with annealed and quenched numerical experiments. 
    Fig.~\ref{fig:annealed-quenched}a shows a comparison between the backtracking DMFT and quenched numerical experiments for the average energy at $N=20,40,80$ and $p=7, c=1,2$ (for $c=2$ we depict the smaller of the energies of the two configurations in the attractor cycle). Larger sizes are prohibitively difficult to sample, as $(p,c)$-backtracking attractors are exponentially rare. We see that the theory is in good agreement with the numerics, and the agreement improves as $N$ increases. 
    Fig.~\ref{fig:annealed-quenched}b shows a comparison between the backtracking DMFT, quenched and annealed numerical experiments for the average energy at $N=20,40,80$ and $p=7, c=1$. In addition to the previous observations, we notice that the annealed and quenched averages seem to start differing at a finite value of $\kgr$, which decreases as $N$ increases.
    Moreover, for $\kgr = 0.025$ (the smallest value considered for $N=80$), we see that the discrepancy between the annealed and quenched numerical experiments get smaller, as well as the discrepancy between both numerical experiments and the theory.
    Fig.~\ref{fig:annealed-quenched}c shows a comparison between the backtracking DMFT, quenched and annealed numerical experiments for the field distribution at $N=80$, $\kgr = 0.1$ and $p=7, c=1$. Again, theory, quenched and annealed experiments are in good agreement.

    We point out that in \cite{dandi2023maximallystable} the authors prove that the $p=0$, $c=1$ quenched entropy is correctly obtained by the annealed computation for a wide range of values of the energy of the fixed points, and of their threshold parameter (in particular, they consider negative $\kgr$, i.e. gapped stable fixed points). We find the same prediction as \cite{dandi2023maximallystable} for the value of $\kgr$ at which $s^{p=0,c=1}_{\rm bk}$ vanishes, as well as the same corresponding energy (see Fig.~\ref{fig:sync-eo-dmft-bk}a-c). 
    On the other hand, in Fig.~\ref{fig:sync-eo-dmft-bk}c we plot the energy of $c=1$ backtracking attractors at negative $\kgr$: we see, for example focusing on $p=1$, that the predicted annealed energy falls below $\egs$ for $\kgr \lessapprox 0.4$, values for which the annealed entropy is still positive. This is a clear contradiction and indication that the annealed approximation becomes incorrect for such negative values of $\kgr$, at least for $p \geq 1$ and $c=1$.

    To summarise, we find that for $\kgr \gtrapprox 0$ the backtracking DMFT seems to be in good quantitative agreement with quenched numerical experiments, the annealed backtracking DMFT may thus be enough to describe the convergence properties of the dynamics in the region of $\kgr \gtrapprox 0$. Clear  discrepancies are observed in the region $\kgr \lessapprox 0$, and for the $c=1$ attractors, suggesting that a quenched analysis is necessary to provide fully solid theoretical analysis of the dynamical phase transition, as well as to predict properties of $c=1$ backtracking attractors at negative threshold (corresponding to transients converging to gapped stable minima).

    Overall, we are puzzled by the fact that the dynamics seems to be converging to the ground state and that the annealed approximation seems to describe the observed behaviour, as the ground state of the SK model lays in a full-RSB phase. On the other hand, one can wonder whether the complex full-RSB correlation structure may arise dynamically in the properties of the dynamics of the DMFT equations. 
    We leave further investigation of this aspect for future work.

    \subsection{The sync-reluctant algorithm}
    
    \begin{figure}
        \centering
        \includegraphics[width=\textwidth]{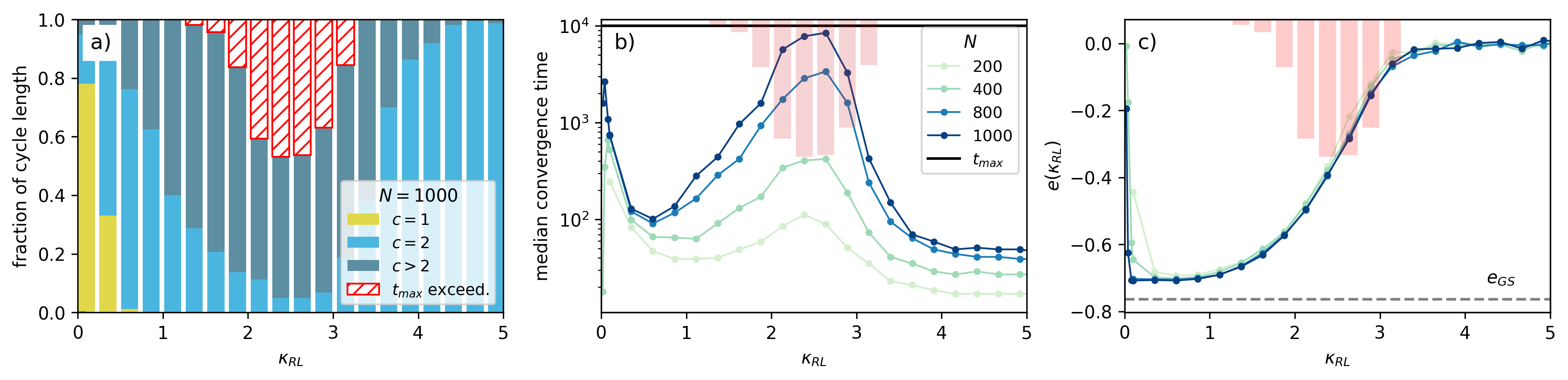}
        \caption{\textbf{Sync-reluctant dynamics -- Numerical experiments.} 
        Numerical experiments for the sync-reluctant dynamics on instances of size $N=[200,400,800,1000]$ for $\krl \in [0,5]$. Each data sample is the average of $160$ runs that were stopped after $t_{max}=10000$ iterations, or at convergence. 
        \textit{(a)} For $N=1000$ we show the fraction of the cycle lengths in the attractors. The shaded red zone denotes all runs for which convergence was not achieved before $t_{max}=10000$ iterations. 
        \textit{(b)} Median transient length until a cyclic attractor is reached as a function of $\krl$ for $N = [200,400,800,1000]$. Red bars show again the fraction of runs where no convergence could be detected until $t_{max}$. 
        \textit{(c)} Median energy after $t_{max}$ iterations as a function of $\krl$ for $N = [200,400,800,1000]$. Red bars show again the fraction of runs where no convergence could be detected until $t_{max}$. 
        }
        \label{fig:sync-rl-empirics}
    \end{figure}

    We now consider a sync-reluctant algorithm, with the dynamical rule $\sigma_{\rm RL}(x, h; \krl) = x \sign(x h (h + \krl x))$.
    This time, we flip only the unsatisfied nodes which have a field \textit{smaller} than $\krl \geq 0$.
    Notice that for $\krl = +\infty$ the sync-greedy dynamics falls back on the sync-Glauber dynamics, while for $\krl \to 0$ the dynamics is frozen, as all initial conditions are fixed points. Moreover, it makes sense to study this dynamics only for $\krl \geq 0$.
    Again, we study to which attractors does this dynamics converge to, what energy to they have, and how long does it take to converge to them.

    \paragraph{Properties of the dynamics.}
    The sync-reluctant dynamics admits \textit{a priori} attractors of all periods $c\geq 1$, contrary to the sync-greedy dynamics. This complicates the use of the backtracking DMFT, as considering $c=1,2$ is not sufficient anymore to provably cover all the phenomenology. 
    It also hinders numerical explorations, as it is more computationally expensive to check for convergence without having a theoretical bound on the size of admissible limit cycles.
    
    \paragraph{Numerical experiments.}
    Fig.~\ref{fig:sync-rl-empirics} sums up the results of the numerical experiments, for sizes up to $N = 1000$, as a function of the threshold parameter $\krl$. Due to the memory cost of checking for arbitrary length cycles, the sizes $N$ for which we run the dynamics are considerably smaller than previously for the sync-greedy dynamics.
    Fig.~\ref{fig:sync-rl-empirics}a shows the fraction of trajectories converging to cycles of length $c=1$, $c=2$, $c>2$ or not yet converged after $t_{\rm max} = 10000$ time steps. We see that in the range $1 < \krl < 3.5$ the majority of the simulated trajectories either did not converge fast enough, or converged to cycles of length $c>2$. In the Appendix, Fig.~\ref{fig:app:cycle-len-hist}, we show a histogram of the cycle lengths for $\krl\cong2.38$, showing that the dynamics can converge to cycles with a broad spectrum of lengths.
    In the same range, the convergence time spikes (Fig.~\ref{fig:sync-rl-empirics}b), compatibly with a linear convergence time $\caO(N)$.
    Fig.~\ref{fig:sync-rl-empirics}c shows the energy reached at convergence as a function of $\krl$. For small but non-zero values of $\krl$ we see that the dynamics reaches energies closer to $\egs$, and this coincides with the appearance of 1-cycles among the dynamical attractors (Fig.~\ref{fig:sync-rl-empirics}a). Still, the energy at convergence seems to plateau at a finite distance from $\egs$.

    \begin{figure}
        \centering
        \includegraphics[width=\textwidth]{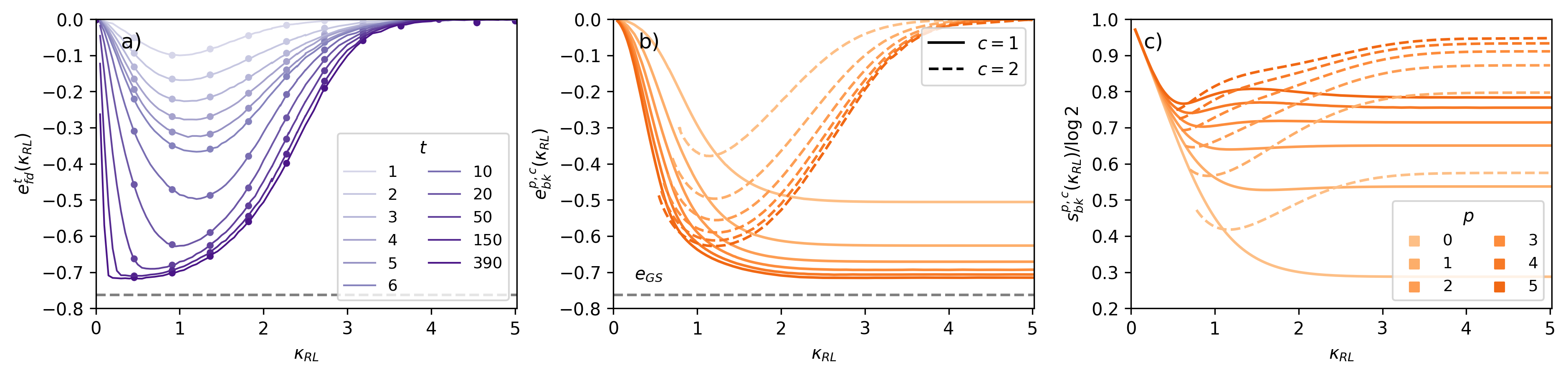}
        \caption{
        \textbf{Sync-reluctant dynamics --- DMFT.}
        \textit{(a)} Energy obtained using the forward DMFT for the sync-reluctant algorithm for various times $t$ between $1$ and $390$. The markers represent the empirical energy on instances of size $N=4000$ after $p$ iterations of the algorithm, where each datapoint is the average over $160$ runs. \textit{(b)} Energy in the attractor obtained using the backtracking DMFT for cycles of length $1$ (solid) and $2$ (dashed) and $p_{BDCM}$ between $0$ and $5$. \textit{(c)} Entropy (normalised by $\log(2)$ from the backtracking DMFT for $c=1$ and $c=2$ backtracking attractors.
        }
        \label{fig:sync-rl-dmft}
    \end{figure}

    \paragraph{Forward DMFT.}
    Forward DMFT (Fig.~\ref{fig:sync-rl-dmft}a) confirms the picture of Fig.~\ref{fig:sync-rl-empirics}a.
    Contrary to the sync-greedy dynamics, we do not observe such a sharp change of behaviour as a function of $\krl$.
    Moreover, we observe that the overall convergence time is much larger, as for the sync-greedy only for $\kgr < 0.1$ the dynamics needed more than $\approx 50$ time steps to converge, while here for all $0 < \krl \lessapprox 3.5$ the dynamics has still not reached convergence after $\approx 400$ time steps. 
    
    \paragraph{Backtracking DMFT.}
    The backtracking DMFT (Fig.~\ref{fig:sync-rl-dmft}b and Fig.~\ref{fig:sync-rl-dmft}c) provides little additional insights. The normalised backtracking entropy $s^{p,c}_{\rm bk}(\krl) / \log(2)$ is $\lessapprox 0.85$ in the regime $1 < \krl < 2.5$ where interesting phenomenology happens, suggesting that we would need to look at backtracking DMFT for much larger values of $p$ to actually predict convergence properties of the dynamics.
    Similar to the sync-greedy case we observe a discontinuous phase transition between convergence to 1-cycles and 2-cycles, but without additional information on the entropies at $c>2$, or results for larger values of transient lengths $p$, the relevance of this transition to the dynamics is not very clear. We tried to solve the backtracking DMFT equations at $c=3,4$, but could not find non-trivial fixed points with entropy larger than the $c=1,2$ case for $p=0, 1$. Larger values of $p$ and $c$ proved too computationally expensive to probe.

    This dynamics highlights the limitations of the backtracking DMFT: if the backtracking entropy does not grow close enough to its maximum value as $p$ increases, backtracking DMFT will surely be describing some rare dynamical motifs, and not the typical convergence structures of the dynamics. Moreover, when long attractors can attract the dynamics, the amount of entropies to be checked increases ($c$ may need to get arbitrarily large), complicating the analysis and providing additional numerical challenges.

    \section{Conclusion}

    In this manuscript, we considered several zero-temperature quenches for the Sherrington-Kirkpatrick Hamiltonian.
    
    We first studied several single-spin flip algorithms already considered in the literature \cite{Parisi1, Coluzzi00, contucci03, Horner07}, providing new numerical simulations for larger systems sizes $N$. We provide updated estimates for the energy at convergence of these algorithms, and for the finite-size scaling exponents. 
    For the particular case of the reluctant dynamics, a kind of ``least greedy'' dynamics, we show that the energy at convergence is compatible with the ground state energy $\egs = -0.763219$, and that the field distribution is pseudo-gapped as predicted analytically for the SK ground state \cite{pankov}.

    To be able to provide an analytical insight, we introduce synchronous dynamical rules inspired by the single spin flips dynamics considered above. These dynamical rules, the sync-greedy and sync-reluctant dynamics, are amenable to analysis through Dynamical Mean Field Theory (DMFT). 
    We study both of them through usual DMFT, and try to characterise their convergence properties using a new variation of DMFT (that we call backtracking DMFT). It is inspired by the Backtracking Dynamical Cavity Method \cite{behrens2023backtracking} and by DMFT for cycle counting \cite{hwang2019number}. We also run extensive numerical experiments.

    The sync-greedy dynamics is the most interesting of the two. We find 
    evidence of a dynamical phase transition between a phase dominated by convergence to zero energy 2-cycles, and a phase dominated by convergence to fixed points with energy close to $\egs$. Close to the phase transition, we find evidence that the energy at convergence may be compatible with $\egs$, albeit the convergence time is largest.
    We highlight that the backtracking DMFT was a crucial tool to characterise the dynamics in this case.
    We, however, note that the data we present cannot refute the possibility that dynamical phase transition between cycles and fixed point does not happen at strictly positive $\kgr > 0$. If this was the case, at least two dynamical phase transitions would exist (symmetrically around $\kgr = 0$), and our extrapolations to $\kgr \to 0$ would need to be corrected. We leave the clarification of this conjecture to future work.

    Surprisingly, we find that the prediction of backtracking DMFT in the annealed approximation seems to agree reasonably with the quenched numerical experiments for a broad range of value of the threshold $\kgr$. We find that understanding whether this observation indeed holds is an important question, that we leave for future work.

    The sync-reluctant dynamics has a less clear phenomenology. It is not clear to which attractors it converges to, and whether there is any value of its threshold parameter for which the energy at convergence is compatible with $\egs$. 
    We remark that in the sync-reluctant case backtracking DMFT was not very helpful, highlighting a limitation of this technique: when convergence is very slow, and cycles of many sizes appear at convergence, backtracking DMFT will struggle to provide information on the convergence properties of algorithms, and will just give statistical information on atypical dynamical motifs.

    Finally, we remark that in Appendix \ref{app.dmft} we derived backtracking DMFT also for non-symmetric couplings, and for more general dynamical rules. It will be interesting to apply this new tool to dynamical processes in neural networks and complex systems, and to potentially discover new properties of their dynamics around attractors. 

\section*{Acknowledgements}

We acknowledge funding from the Swiss National Science Foundation grants TMPFP2 210012.

We would like to thank Emanuele Troiani for providing useful tips on the DMFT numerics, Pierfrancesco Urbani for suggesting to look at the pseudo-gap of the field distribution, Guy Bresler for suggesting to look at the statistics of flips per variable and Ahmed El Alaoui for providing an implementation of IAMP algorithm. 
We thank Giulio Biroli, Alejandro Lage Castellanos, and Guilhem Semerjian for enlightening discussions.  
    
    \bibliography{biblio}

    \appendix

\newpage

    \section{Single Spin Flip Dynamics: Supporting Numerics}
    \label{app:numerical_e_est}
    
\begin{figure}[!ht]
        \centering
        \includegraphics[width=\textwidth]{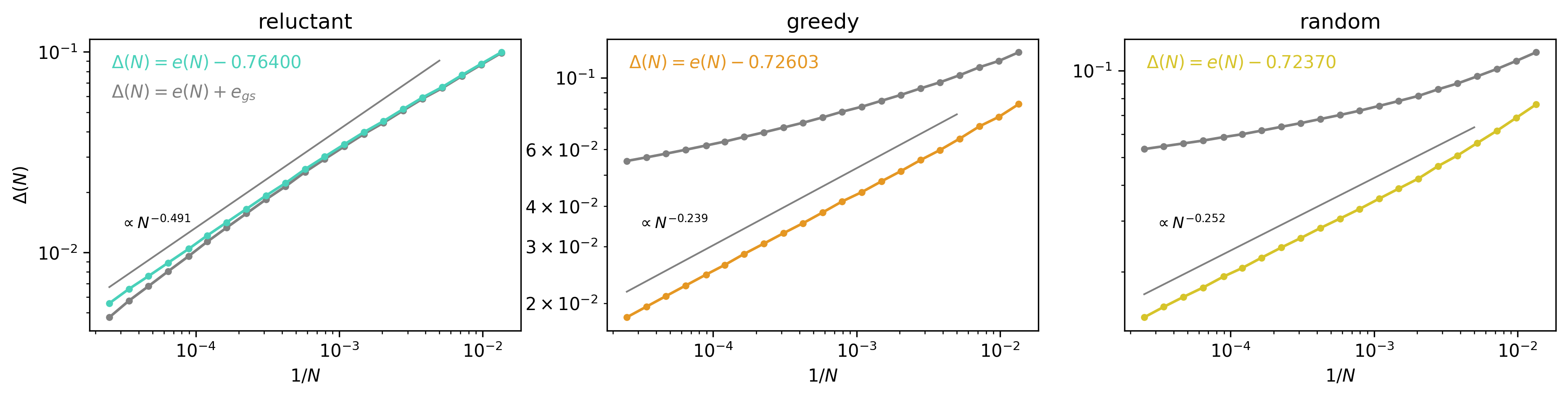}
        \caption{\textbf{Fits of the energy as a function of $N$ for the single spin flip algorithm.} In this plot we show how the distance to a given energy $\Delta(N)$ evolves for the reluctant, greedy and random algorithm. For each algorithm we compare the best fit, derived from Fig.~\ref{fig:e_conv_vs_n} with the distance to the ground state $e_{GS}$. The data stems from Fig.~\ref{fig:e_conv_vs_n}. The exponents of the displayed slopes were obtained by fitting the function $\Delta(N) = b N^{-c} + E_{est}$ to the same points from which we obtained $E_{est}$ itself in Fig.~\ref{fig:e_conv_vs_n}.}
        \label{fig:app:approximate-fits}
    \end{figure}

    In Fig.~\ref{fig:app:approximate-fits} we show the fits that we derived in Fig.~\ref{fig:e_conv_vs_n}b by selecting the intercept that gives the fit with the smallest curvature. For comparison, we also plot the fit to the ground state energy in grey, showing that the intercept $e_{\rm est}$ estimated for the reluctant dynamics is very close to $\egs$, while for the greedy and random single spin flip algorithms there is a noticeable difference.

\begin{figure}[!htb]
        \centering
        \includegraphics[width=\textwidth]{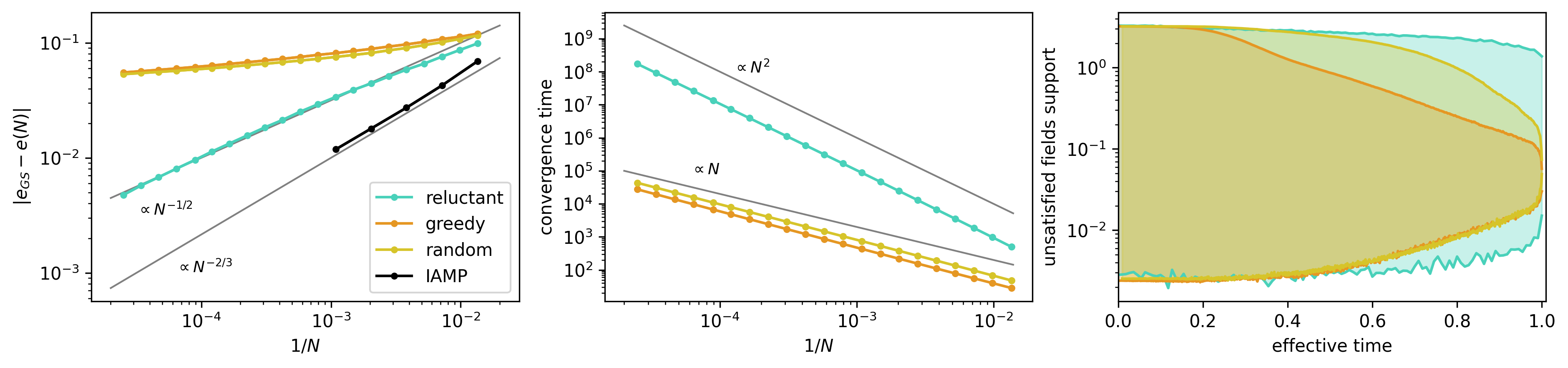}
        \caption{\textbf{Energy, convergence behaviour and field support for single spin flip dynamics.} 
        \textit{(a)} Energy difference to the ground state $\egs$ at convergence (information on the data can be found in the caption of Fig.~\ref{fig:e_conv_vs_n}). We see that the reluctant and the IAMP algorithms are compatible with an energy at convergence in the thermodynamic limit equal to $\egs$.
        \textit{(b)} The convergence speed of reluctant is of order $\caO(N^2)$, while greedy and random converge faster in $\caO(N)$ time steps. 
        \textit{(c)} Support of the absolute value of the unsatisfied fields $|x_ih_i|$ with $x_ih_i < 0$ with $N=1707$. The effective time differs for each algorithm as a function of the size $n$:  $t_{\mathrm{eff}}^{\mathrm{reluctant}}= \frac{t}{.11N^2}$,$t_{\mathrm{eff}}^{\mathrm{greedy}}= \frac{t}{.63N}$,$t_{\mathrm{eff}}^{\mathrm{random}}= \frac{t}{1.05N}$. Every trajectory is the average of 1000 samples (except for the reluctant algorithm, where we sampled 29 trajectories). To average, we align the trajectories at their convergence, so that the last iteration for any sample is aligned at effective time $t_{\mathrm{eff}} = 1$. Since different samples have different convergence times, the start may also be in the negative time regime.
        \textit{(d)} Distribution of local fields at convergence. We see no gap at $h = 0$.
        }
        \label{fig:app:sequential-algorithms-additional}
    \end{figure}

\begin{figure}[!ht]
        \centering
        \includegraphics[width=\textwidth]{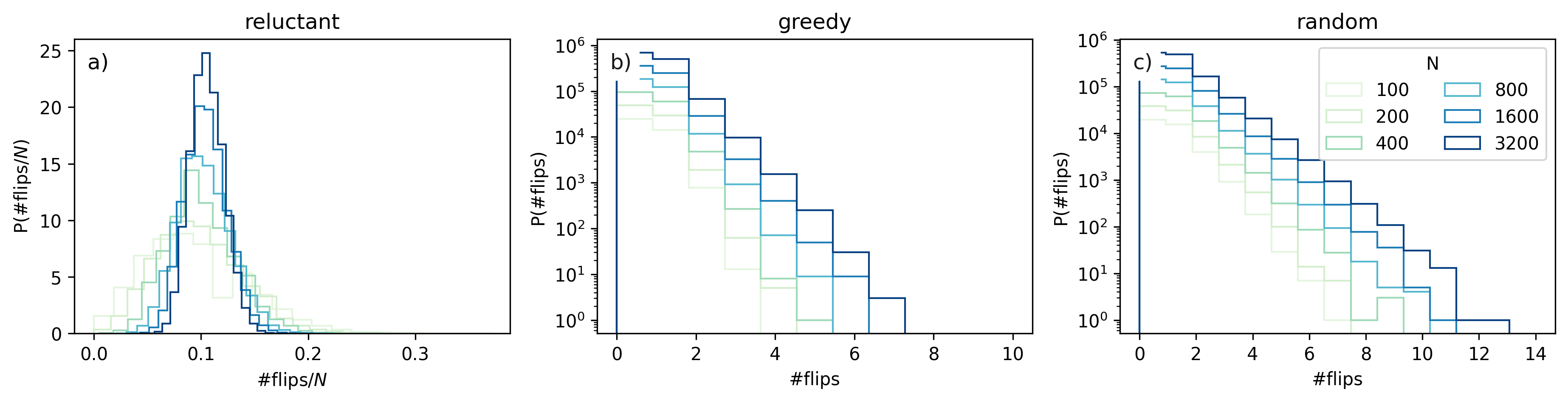}
        \includegraphics[width=0.66\textwidth]{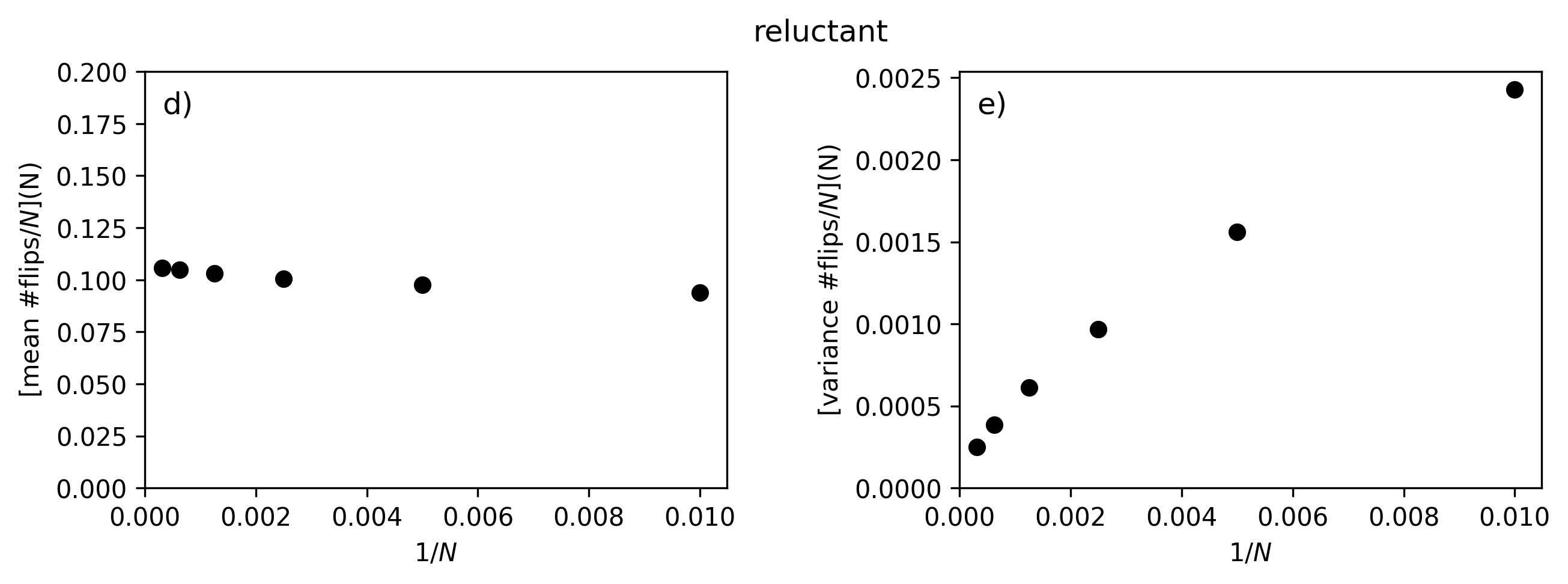}
        \caption{\textbf{Numerics on the flips per variable during a single run of single spin flip dynamics.}
        \textit{(a,b,c)} Histograms of the number of flips per spin during a run of a single algorithm (reluctant, greedy, random) for $N=\{100,200,400,800,1600,3200\}$. Each algorithm was run 400 times to extract statistics over $N$ variables each,  every sample was a new random configuration and graph structure. 
        Note that for the reluctant algorithm we normalize the number of flips per node by $1/N$.
        \textit{(d)} Mean of the distribution of the flips for the reluctant algorithm in (a) as a function of $N$. \textit{(e)} Variance of the reluctant algorithm in (a) as a function of $N$.
        }
        \label{fig:app:flip-counts}
    \end{figure}

    In Fig.~\ref{fig:app:sequential-algorithms-additional} we provide additional information on the convergence behaviour for all the three algorithms. We show the convergence behaviour in the energy density, and compare with IAMP \cite{alaoui2020algorithmic}, which finds lower energies for all sizes $n$ that we tested than the single spin flip algorithms.
    Fig.~\ref{fig:app:sequential-algorithms-additional}b shows that the reluctant algorithm has convergence in roughly $O(N^2)$, while greedy and random take $O(N)$ time, i.e. are faster.
    Finally, we examine the support of the unsatisfied fields $|h_i|$ in Fig.~\ref{fig:app:sequential-algorithms-additional}c over the effective runtime of each algorithm, showing that the distributions differ significantly. This shows  that even though the convergence time and energy are similar for the greedy and random single spin flip algorithms, their effects on the field distribution is qualitatively different over time.
    In Fig.~\ref{fig:app:flip-counts} we show that the number of flips per spin is also qualitatively different. For the reluctant the flips per spin are a function of the graph size $N$. Contrarily, for the greedy and random algorithms we observe that many spins are not flipped at all, while a few are flipped often, decaying as a power law.
    Further, for the reluctant algorithm the distribution, centered around roughly $0.11N$, seems to approach a delta function, as the variance shrinks as $N$ grows.

\newpage

    \section{Synchronous Dynamics: Supporting Numerics}

\begin{figure}[!htb]
        \centering
        \includegraphics[width=0.33\textwidth]{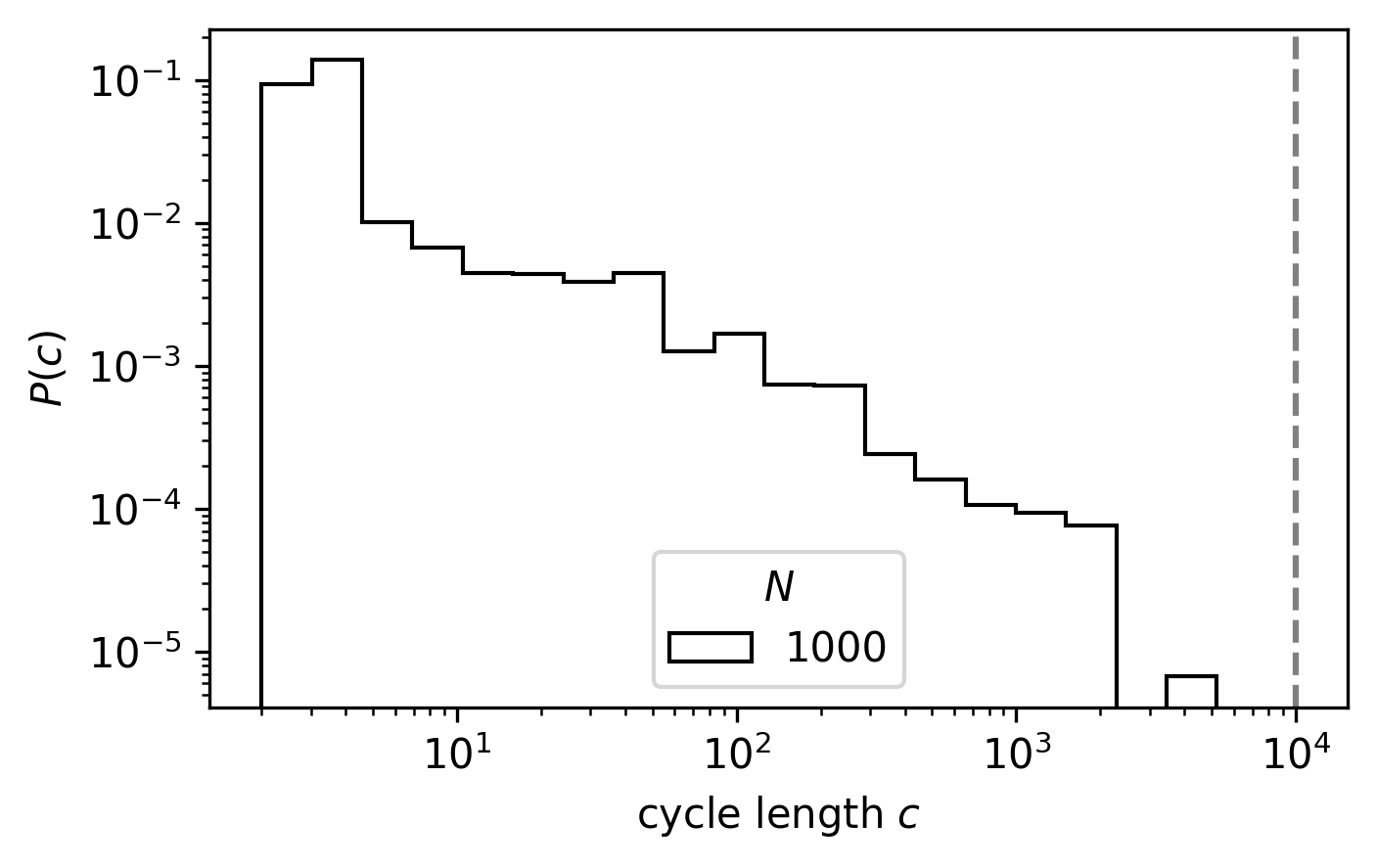}
        \caption{\textbf{Cycle length distribution for the sync-reluctant dynamics.} From the data in Fig.~\ref{fig:sync-rl-empirics} we select $\kappa=2.3846$ and show the distribution of cycles in a log scale. The data includes only trajectories where we were able to detect the attractor within $t=10,000$ times steps. Therefore, only $\sim 53.13$\% of the 160 samples are shown. We see that cycles of lengths up to $\approx 10^3$ where found.}
        \label{fig:app:cycle-len-hist}
    \end{figure}

    In Figure~\ref{fig:app:cycle-len-hist} we show the distribution of the cycle lengths for a fixed value of $\kappa_{RL}$. 

\begin{figure}[!htb]
        \centering
        \includegraphics[width=0.33\textwidth]{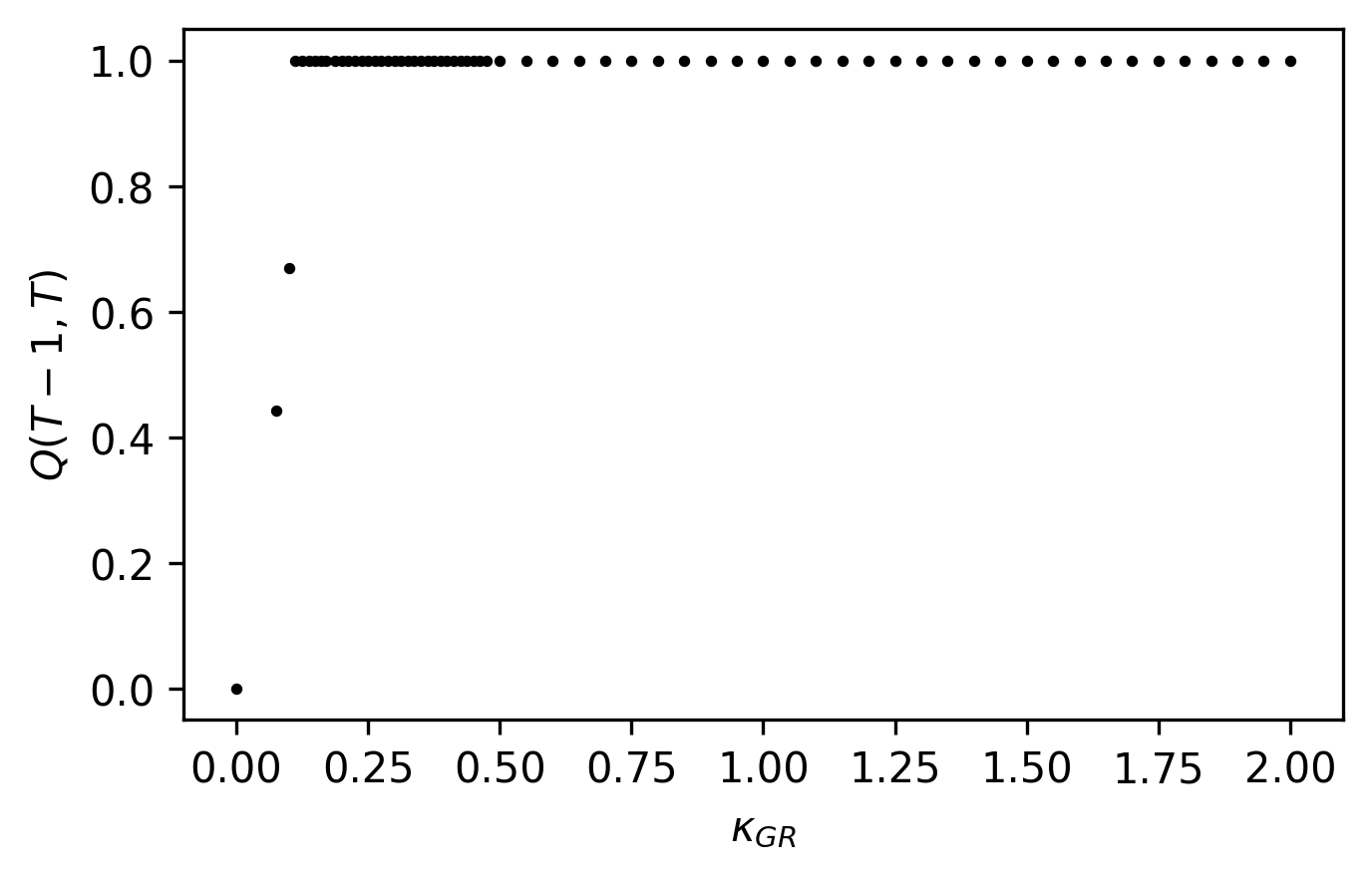}
        \caption{\textbf{Correlation $Q(T-1,T)$ for the forward dynamics of the sync-greedy algorithm.} From the same data plotted in Fig.~\ref{fig:sync-eo-dmft-fd}, we compute $Q(T-1, T)$ as a function of $\kgr$. We see that for $\kgr \geq 0.1$, where the forward DMFT reaches convergence, $Q(T-1,T) \approx 1)$, meaning that the dynamics converges to fixed points.}
        \label{fig:app:correlationfd}
    \end{figure}

    In Fig.~\ref{fig:app:correlationfd} we show the numerical results for $Q(T-1,T)$ from the forward DMFT. We see that for $\kgr \geq 0.1$, where the forward DMFT reaches convergence, $Q(T-1,T) \approx 1)$, meaning that the dynamics converges to fixed points.
    
    \section{Single Spin Flip Dynamics on the 3-spin Model}\label{ap:3spin}

    The greedy and reluctant single spin flip dynamics generalize naturally to $p$-spin models, where $p$ instead of $2$ variables interact through a symmetric Gaussian interaction term.

\begin{figure}
        \centering
        \includegraphics[width=\textwidth]{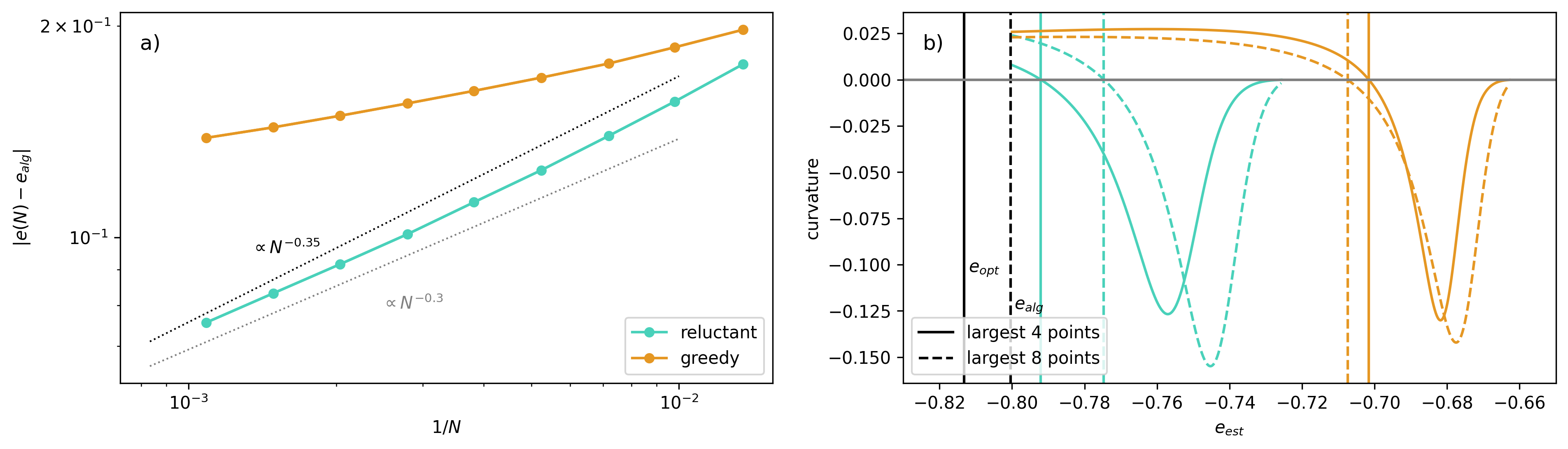}
        \caption{The reluctant and greedy algorithm for the $3$-spin model. We use systems of size $N \in \{1261,920,672,491,358,262,191,139\}$. \textit{(Left)} Difference of the average energy at convergence $e(N)$ of $16,000$ initializations of the disorder $J$ for every $N$ with the algorithmic threshold $e_{\rm alg}$ \cite{alaoui2020algorithmic}. \textit{(Right)} Curvature of a quadratic fit of the 8 (4) points with the largest $N$ on a log-log plot for the distance for different possible  of the energy $e_{est}$, similar to Figure~\ref{fig:e_conv_vs_n}, right. We observe there is a high variance in the fit. For the reluctant algorithm, the fit on the last 4 points is better  compatible with $e_{\rm alg}$ than the last 8. For the greedy algorithm, the difference of the prediction is much smaller than the distance to $e_{\rm alg}$, thus effectively excluding the possibility that the greedy algorithm reaches $e_{\rm alg}$.}
        \label{fig:pspin}
    \end{figure}

    In the following we consider the two algorithms for the $3$-spin model, i.e. $x \in \pm1^N$, $J_{ijk}$ are again i.i.d. standard Gaussian and where the Hamiltonian and local fields are defined as
    \begin{align}
    h_i &= \frac{\sqrt{3}}{N}\sum_{1 \leq j < k \leq N} J_{ijk} x_j x_k\,, \\
        H^{3\mathrm{-spin}}_J(x) &= -\frac{\sqrt{3}}{N} \sum_{1 \leq i < j < k \leq N} J_{ijk} x_i x_j x_k = -  \sum_{1 \leq i \leq N} x_i h_i\,\\
        e(N) &= - \frac{1}{N}  \sum_{1 \leq i \leq N} x_i h_i\, .
    \end{align}
With this definition of the fields, we can define the reluctant and greedy dynamics in analogy to Section~\ref{sec.single}.
In Figure~\ref{fig:pspin} we show simulations of both the greedy and reluctant algorithm up to $N=1726$.
Since each iteration, acting on a three-dimensional tensor, has a higher complexity, we are limited to this size.
There are known algorithms \cite{alaoui2020algorithmic} that provably converge to the energy density of the so-called \textit{algorithmic threshold}, $e_{\rm alg} \sim - 0.8004$, but not the ground state energy density $e_{\rm opt} \sim - 0.8135$. In fact there are lower bounds \cite{huang2022tight} suggesting strongly that $e_{\rm alg}$ is the lowest energy reachable efficiently. 

The numerical results are inconclusive on whether the reluctant dynamics reaches the algorithmic threshold or not but do not exclude that possibility either. For the greedy algorithm, it is rather clear from the data that they do not reach $e_{\rm alg}$.  Larger system sizes or analytical studies are needed to clarify this point.

% Both algorithms seem in principle compatible with going to the algorithmically reachable energy. 
% However being limited to a comparably small size of $N$ we take our simulations to be inconclusive, as to whether one of the two processes arrives at the algorithmic energy threshold.

    \section{Derivation of the DMFT equations for synchronous dynamics}\label{app.dmft}

    In this section, we study the class of dynamical processes defined by the update rule
    \begin{equation}
        x_i(t+1) = \sigma\left(x_i(t) , h_i(t) \right)
    \end{equation}
    where $\sigma : \{ \pm 1 \} \times \bbR \to \{ \pm 1 \}$ is a generic update function. 
    We consider synchronous updates, i.e. all spins $\{x_i\}_{i=1}^N$ are updated at the same time.

    We will consider the coupling matrix given by
    \begin{equation}\label{def.J}
        J_{ij} = \left(1 - \epsilon \right) S_{ij} + \epsilon A_{ij}
    \end{equation}
    where $S$ and $A$ are random matrices with i.i.d. standard Gaussian entries, respectively symmetric and antisymmetric, both with zero diagonal.
    In the main text we focus on the symmetric case $\epsilon = 0$, but in this Appendix we perform computations for generic $0 \leq \epsilon \leq 1$.
    For $\epsilon = 0$ the interaction is symmetric, for $\epsilon = 1$ is antisymmetric, and for $\epsilon = 1/2$ is a Gaussian matrix with iid entries, as the only possible non-zero correlation vanishes at $\epsilon = 1/2$:
    \begin{equation}
        \begin{split}
            \EE[J_{ij} J_{ji}] 
            &= 
            \left(1 - \epsilon \right)^2 \EE[ S_{ij} S_{ji} ]
            +
            \epsilon^2 \EE[ A_{ij} A_{ji} ]
            =
            \left(1 - \epsilon \right)^2 -
            \epsilon^2 \, .
        \end{split}
    \end{equation}
    Notice that in this case $J_{ii} \neq 0$, contrary to our assumption in the main text. This is not a problem, as the diagonal element contribute only to subleading orders of the observables.

    In the following, indices $i, j, k, \dots$ denote spins and run from $1$ to $N$, indices $t, s, u \dots$ denote times and run from $1$ to $T$.
    We denote by $x_i(t)$ the value of the $i$-th spin at time $t$, by $x(t) = \{x_1(t), \dots, x_N(t)\}$ the collective value of all spins at time $t$, by $\ux_i = \{x_i(1), \dots, x_i(T+1)\}$ the value of the $i$-th spin at all times and finally by $\ux$ the value of all spins at all times.
    We will often drop the index $i$ from $\ux_i$ whenever it is clear from the context that just one spin is considered, instead of the full system.

    \subsection{Preliminaries}

    We consider the following partition function
    \begin{equation}
        Z(\caT, J) = \sum_{\substack{\ux_1 \in \caT}}\sum_{\substack{\ux_2 \in \caT}}   \dots \sum_{\substack{\ux_N \in \caT}} \prod_{i=1}^N \prod_{t=1}^{T} 
        \delta\left[x_i(t+1) = \sigma\left( x_i(t), h_i(t) \right)\right]
    \end{equation}
    where $J$ is the coupling matrix defined in \eqref{def.J}, $\delta[x=y]$ is the Kronecker delta and $\caT \subseteq \{\pm 1\}^{T+1}$ is a set of allowed single-spin trajectories.
    We also define the annealed and quenched entropies as
    \begin{equation}
        s_{\rm ann}(\caT) = \lim_{N \to +\infty} \frac{1}{N} \log \EE_J Z(\caT, J)
        \mathand
        s_{\rm quench}(\caT) = \lim_{N \to +\infty} \frac{1}{N} \EE_J \log Z(\caT, J) \, .
    \end{equation}

    This partition function counts how many initial conditions $x(1)$ lead to a (deterministic) trajectory $\ux = (x(1), x(2), \dots, x(T+1))$ such that $\ux_i \in \caT$ for all spins $x_i$.
    To be concrete, we will be interested in the following cases:
    \begin{itemize}
        \item "forward dynamics", where $\caT = \caT_{\rm fd} = \{\pm 1\}^{T+1}$. All trajectories are allowed.
        \item "$(p,c)$-backtracking dynamics", where $\caT = \caT_{\rm bk}(p,c) = \{\ux = \{\pm 1\}^{T+1} \,\,\text{such that}\,\, x(p+1) = x(p+c+1)\}$. Allowed trajectories are transients of length $p$ ending into a cycle of length $c$ \cite{behrens2023backtracking}.
    \end{itemize}

    In the forward dynamics case, the fact that the dynamics is deterministic readily implies that $Z(\caT_{\rm fd}, J) = 2^N$ for all $J$. Computing the partition function is still interesting, as we expect it to concentrate providing us with the value of certain order parameters characterising the average dynamics. 
    Notice also that the fact that the partition function is constant implies that 
    \begin{equation}
        s_{\rm ann}(\caT) = s_{\rm quench}(\caT) = \log 2 \, ,
    \end{equation}
    i.e. annealed and quenched averages coincide. 
    Thus, in the forward dynamics case, an annealed average suffices to obtain the correct values for the typical properties of the dynamics.

    In the $(p,c)$-backtracking dynamics case, the partition function is not constant anymore. Thus, \textit{a priori} annealed and quenched averages do not coincide, and the typical properties of the dynamics should be studied using the quenched entropy. In this work, we consider the annealed entropy (an upper bound to the quenched entropy), and verify numerically that it provides reliable predictions for the typical properties of the dynamics.

    Finally, we consider the slightly more generic partition function
    \begin{equation}
        Z(\caT, J, g) = 
        \sum_{\substack{\ux_1 \in \caT}}\sum_{\substack{\ux_2 \in \caT}}   \dots \sum_{\substack{\ux_N \in \caT}} \prod_{i=1}^N e^{g(\ux_i, \uh_i(\ux))}
        \prod_{t=1}^{T} 
        \delta\left[x_i(t+1) = \sigma\left( x_i(t), h_i(t) \right)\right]
    \end{equation}
    where the function $g$ may contain any spin-factorised observable (i.e. observable of the form $\sum_i \dots$), which can be useful to keep track of the average value of the energy, and the dynamical correlators.
    We will keep $g$ generic in the computations, but in our minds it has the form
    \begin{equation}
        g(\ux_i, \uh_i(\ux)) = 
        \sum_{t < s} \omega^{xx}_{t,s} x_i(t) x_i(s)
        + \sum_{t,s} \omega^{xh}_{t,s} x_i(t) h_i(s)
        + \sum_{t \leq s} \omega^{hh}_{t,s} h_i(t) h_i(s)
    \end{equation}
    allowing us to keep track of the dynamical correlators
    \begin{equation}
        \begin{split}
            C^{xx}(t,s) &= \frac{1}{N} \sum_i x_i(t) x_i(s) 
            \sim \del_{\omega^{xx}_{t,s}} \left[ \frac{1}{N} \EE_J \log Z(\caT, J, g) \right]_{\omega = 0}
            \, ,\\
            C^{xh}(t,s) &= \frac{1}{N} \sum_i x_i(t) h_i(s) 
            \sim \del_{\omega^{xh}_{t,s}} \left[ \frac{1}{N} \EE_J \log Z(\caT, J, g) \right]_{\omega = 0}
            \, ,\\
            C^{hh}(t,s) &= \frac{1}{N} \sum_i h_i(t) h_i(s) 
            \sim \del_{\omega^{hh}_{t,s}} \left[ \frac{1}{N} \EE_J \log Z(\caT, J, g) \right]_{\omega = 0}
            \, , 
        \end{split}
    \end{equation}
    including the energy of the SK model (maningful only for $J$ symmetric)
    \begin{equation}
        E(t) = - \frac{1}{2} C^{xh}(t,t) \, .
    \end{equation} 
    Setting the "temperatures" $\omega \neq 0$ can also allow to condition on the dynamical correlators to have specific values, as it is usually the case in standard statistical mechanics when the temperature is tuned to fix the desired internal energy.

    \subsection{Annealed disorder average}

    We loosely follow \cite{opper94,hwang2019number} for this derivation, extending it to more generic dynamical rules (they only consider the sync-Glauber dynamics $\sigma(x,h) = \sign(h)$) and more generic dynamical constraints (\cite{opper94} considers only the forward dynamics, \cite{hwang2019number} considers also the $(p,c)$-backtracking dynamics with $p=0$).

    We compute the annealed average $\EE_J Z$. Defining the shorthand
    \begin{equation}
        \theta_{\caT, \sigma}(\ux_i, \uh_i(\ux)) 
        =
        \delta(\ux_i \in \caT)
        \prod_{t=1}^{T} 
        \delta\Big[x_i(t+1) = \sigma\big( x_i(t), h_i(x(t)) \big) \Big] \, ,
    \end{equation}
    we have
    \begin{equation}
        \begin{split}
            \EE_J Z(\caT, J)
            &= 
            \EE_J
            \sum_{\ux} 
            \prod_{i} 
            \theta_{\caT, \sigma}(\ux_i, \uh_i(\ux)) 
            e^{g(\ux_i, \uh_i(\ux))}
            \\
            &= 
            (2\pi)^{-TN}
            \sum_{\ux} 
            \int d\uh \,
            \bigg[ \prod_i \theta_{\caT, \sigma}(\ux_i, \uh_i) e^{g(\ux_i, \uh_i)}\bigg]
            \int d\umu \,
            \EE_J
            \exp 
            \bigg[
                i \sum_{i, t} h_i^t \mu_i^t
                - \frac{i}{\sqrt{N}} \sum_{i,j,t} J_{ij} \mu_i^t x_j^t
            \bigg]
        \end{split}
    \end{equation}
    where we used a delta function (written in its Fourier representation) to isolate the local field $h_i(t)$ (for times $t=1, \dots T$), as it is the only quantity that depends on the disorder.
    We now perform the disorder average. 
    \begin{equation}
        \begin{split}
            \EE_J &
            \exp 
            \bigg[
                - \frac{i}{\sqrt{N}} \sum_{i,j,t} J_{ij} \mu_i^t x_j^t
            \bigg]
            =
            \EE_{S,A}
            \exp 
            \bigg[
                - \frac{i (1-\epsilon)}{\sqrt{N}} \sum_{i,j,t} S_{ij} \mu_i^t x_j^t
                - \frac{i \epsilon}{\sqrt{N}} \sum_{i,j,t} A_{ij} \mu_i^t x_j^t
            \bigg]
            \\
            &=
            \EE_{S,A}
            \exp 
            \bigg[
                - \frac{i (1-\epsilon)}{\sqrt{N}} 
                \sum_{i < j} S_{ij} 
                \big( \sum_t \mu_i^t x_j^t + \sum_s \mu_j^s x_i^s \big)
                - \frac{i \epsilon}{\sqrt{N}} 
                \sum_{i < j} A_{ij} 
                \big( \sum_t \mu_i^t x_j^t - \sum_s \mu_j^s x_i^s  \big)
            \bigg]
            \\
            &=
            \exp 
            \bigg[
                - \frac{(1-\epsilon)^2}{2N} 
                \sum_{i < j} 
                \big( \sum_t \mu_i^t x_j^t + \sum_s \mu_j^s x_i^s \big)^2
                - \frac{\epsilon^2}{2N} 
                \sum_{i < j} 
                \big( \sum_t \mu_i^t x_j^t - \sum_s \mu_j^s x_i^s  \big)^2
            \bigg]
            \\
            &\approx
            \exp 
            \bigg[
                - \frac{(1-\epsilon)^2}{4 N} 
                \sum_{i, j} 
                \big( \sum_t \mu_i^t x_j^t + \sum_s \mu_j^s x_i^s \big)^2
                - \frac{\epsilon^2}{4 N} 
                \sum_{i, j} 
                \big( \sum_t \mu_i^t x_j^t - \sum_s \mu_j^s x_i^s  \big)^2
            \bigg]
            \\
            &\approx
            \exp 
            \bigg[
                - \frac{(1-\epsilon)^2 + \epsilon^2}{2N} 
                \sum_{i, j} 
                \big( \sum_t \mu_i^t x_j^t \big)^2
                - \frac{(1-\epsilon)^2 - \epsilon^2}{2N} 
                \sum_{i, j} 
                \big( \sum_t \mu_i^t x_j^t \big)\big( \sum_s \mu_j^s x_i^s \big)
            \bigg]
        \end{split}
    \end{equation}
    where we did not track explicitly the term $i=j$ in the first sum as it is subleading.
Finally
\begin{equation}
    \begin{split}
        \EE_J &
        \exp 
        \bigg[
            - \frac{i}{\sqrt{N}} \sum_{i,j,t} J_{ij} \mu_i^t x_j^t
        \bigg]
        \approx
        \exp N \nu_\epsilon
        \bigg[
            \sum_{t < s} Q(t,s) R(t,s)
            + \frac{1}{2} \sum_{t} R(t,t)
            + \frac{\eta_\epsilon}{2} \sum_{t,s} S(t,s) S(s,t)
        \bigg]
    \end{split}
\end{equation}
where we introduced the order parameters
\begin{equation}
    \begin{split}
        N Q(t,s) &= \sum_i x_i^t x_i^s \mathfor t < s \, ,\\
        N R(t,s) &= - \sum_i \mu_i^t \mu_i^s \mathfor t \leq s \, ,\\
        N S(t,s) &= - i \sum_i x_i^t \mu_i^s \mathfor t, s \, 
    \end{split}
\end{equation}
(the rationale for the signs is to always write observables as a function of $i \mu$)
and defined
\begin{equation}
    \eta_\epsilon = \frac{(1-\epsilon)^2 - \epsilon^2}{(1-\epsilon)^2 + \epsilon^2}
    \mathand
    \nu_\epsilon = (1-\epsilon)^2 + \epsilon^2 > 0\, .
\end{equation}
Notice that while the current definition of the order parameters is the natural one, we will later find it convenient to slightly change the definitions, see \eqref{op.ch1} and \eqref{op.ch2}.
Also notice that for $\epsilon = 0$ (symmetric couplings), $\eta_\epsilon = \nu_\epsilon = 1$.

Using Dirac's deltas (in Fourier representation) to enforce the definition of the order parameters gives (discarding all non-leading contributions)
\begin{equation}
    \begin{split}
        \EE_J &
        \exp 
        \bigg[
            - \frac{i}{\sqrt{N}} \sum_{i,j,t} J_{ij} \mu_i^t x_j^t
        \bigg]
        \\
        &\approx
        \int \text{op.}
        \exp N \nu_\epsilon
        \bigg[
            \sum_{t < s} Q(t,s) R(t,s)
            + \frac{1}{2} \sum_{t} R(t,t)
            + \frac{1}{2} \eta_\epsilon \sum_{t,s} S(t,s) S(s,t)
        \bigg]
        \\ 
        &\quad\times
        \exp \nu_\epsilon \sum_{t < s}
        \bigg[
            i N \hQ(t,s) Q(t,s) 
            - i \hQ(t,s) \sum_i x_i^t x_i^s
        \bigg]
        \\ 
        &\quad\times
        \exp \nu_\epsilon \sum_{t \leq s}
        \bigg[
            i N \hR(t,s) R(t,s) 
            + i \hR(t,s) \sum_i \mu_i^t \mu_i^s
        \bigg]
        \\ 
        &\quad\times
        \exp \nu_\epsilon \sum_{t , s}
        \bigg[
            i N \hS(t,s) S(t,s) 
            - \hS(t,s) \sum_i x_i^t \mu_i^s
        \bigg]
        \\
        &\approx
        \int \text{op.}
        \exp N \nu_\epsilon
        \bigg[
            \sum_{t < s} Q(t,s) R(t,s)
            + \frac{1}{2} \sum_{t} R(t,t)
            + \frac{1}{2} \eta_\epsilon \sum_{t,s} S(t,s) S(s,t)
        \bigg]
        \\ 
        &\quad\times
        \exp N \nu_\epsilon 
        \bigg[
            i \sum_{t < s}\hQ(t,s) Q(t,s) 
            + i \sum_{t \leq s}\hR(t,s) R(t,s) 
            + i \sum_{t ,s}\hS(t,s) S(t,s) 
        \bigg]
        \\ 
        &\quad\times
        \prod_i
        \exp \nu_\epsilon 
        \bigg[
            i \sum_{t \leq s} \hR(t,s) \mu_i^t \mu_i^s
            - i \sum_{t < s} \hQ(t,s) x_i^t x_i^s
            - \sum_{t, s} \hS(t,s) x_i^t \mu_i^s
        \bigg]
    \end{split}
\end{equation}
where $\int \text{op.}$ stands for integration over all order parameters $Q,R,S$ and their conjugate variables $\hQ, \hR, \hS$.
We now see that also the disorder-dependent part of the parition function factorises over $i$, similarly to what happend to the disorder-independent part.
Thus, the partition function can be rewritten as (using the fact that $i$ is now a dummy index)
\begin{equation}
    \begin{split}
        Z &\approx
        \int \text{op.}
        \exp N \nu_\epsilon
        \bigg[
            \sum_{t < s} Q(t,s) R(t,s)
            + \frac{1}{2} \sum_{t} R(t,t)
            + \frac{1}{2} \eta_\epsilon \sum_{t,s} S(t,s) S(s,t)
        \bigg]
        \\ 
        &\quad\times
        \exp N \nu_\epsilon 
        \bigg[
            i \sum_{t < s}\hQ(t,s) Q(t,s) 
            + i \sum_{t \leq s}\hR(t,s) R(t,s) 
            + i \sum_{t ,s}\hS(t,s) S(t,s) 
        \bigg]
        \\ 
        &\quad\times
        \exp N \log
        \Bigg[
            \frac{1}{(2\pi)^{T}}
            \sum_{\ux} 
            \int d\uh \,
            \theta_{\caT, \sigma}(\ux, \uh) e^{
                g(\ux, \uh)
                - i \nu_\epsilon \sum_{t < s} \hQ(t,s) x^t x^s
            }
            \\
            &\quad\quad\quad\quad\times
            \int d\umu \,
            \exp 
            \bigg[
                i \sum_{t} h^t \mu^t
                + i \nu_\epsilon 
                \sum_{t \leq s} 
                \hR(t,s) \mu^t \mu^s
                - \nu_\epsilon  
                \sum_{t, s} 
                \hS(t,s) x^t \mu^s
            \bigg]
        \Bigg]
    \end{split}
\end{equation}
where now $\ux$ denotes just a $(T+1)$-long trajectory, and not one per spin as before, and similarly $\uh$ and $\umu$ are single-spin quantities.

We notice now that the integral in $\umu$ is Gaussian. Indeed
\begin{equation}
    \begin{split}
        \frac{1}{(2\pi)^{T}}
        \int & d\umu \, 
            \exp 
            \bigg[
                i \sum_{t} h^t \mu^t
                + i \nu_\epsilon 
                \sum_{t \leq s} 
                \hR(t,s) \mu^t \mu^s
                - \nu_\epsilon  
                \sum_{t, s} 
                \hS(t,s) x^t \mu^s
            \bigg]
        \\
        &=
        \frac{1}{(2\pi)^{T}}
        \int d\umu \,
            \exp 
            \bigg[
                i \sum_{s} \mu^s
                \big(
                    h^s 
                    + i \nu_\epsilon \sum_{t} 
                    \hS(t,s) x^t 
                \big)
                + i \nu_\epsilon 
                \sum_{t} 
                \hR(t,t) \mu^t \mu^t
                + i \frac{\nu_\epsilon}{2}
                \sum_{t \neq s} 
                \hR(t,s) \mu^t \mu^s
            \bigg]
        \\
        &=
        \frac{1}{\sqrt{(2\pi)^{T} \det A}}
        \exp
        \bigg[
            - \frac{1}{2}
            \sum_{t,s}
            \big(
                h^t + i \nu_\epsilon \sum_{u} 
                \hS(u,t) x^u
            \big)
            A^{-1}(t,s)
            \big(
                h^s + i \nu_\epsilon \sum_{w} 
                \hS(w,s) x^w
            \big)
        \bigg]
        \\
        &=
        \caN\bigg( \uh ; - i \nu_\epsilon \sum_{w} \hS(w,s) x^w ; 
            - i \nu_\epsilon \big(
            \delta(t=s) 
            2 \hR(t,t)
            + \delta(t\neq s) \hR(t,s)
        \big)\bigg)
    \end{split}
\end{equation}
where we called 
\begin{equation}
    A(t,s) 
    = - i \nu_\epsilon \bigg(
        \delta(t=s) 
        2 \hR(t,t)
        + \delta(t\neq s) \hR(t,s)
    \bigg)
\end{equation}
and $\caN(a,b,c)$ is the normalised multivariate Gaussian density for the variable $a$ with mean $b$ and covariance $c$.
\begin{equation}
    \begin{split}
        Z &\approx
        \int \text{op.}
        \exp N \nu_\epsilon
        \bigg[
            \sum_{t < s} Q(t,s) R(t,s)
            + \frac{1}{2} \sum_{t} R(t,t)
            + \frac{1}{2} \eta_\epsilon \sum_{t,s} S(t,s) S(s,t)
        \bigg]
        \\ 
        &\quad\quad\quad\times
        \exp N \nu_\epsilon 
        \bigg[
            i \sum_{t < s}\hQ(t,s) Q(t,s) 
            + i \sum_{t \leq s}\hR(t,s) R(t,s) 
            + i \sum_{t ,s}\hS(t,s) S(t,s) 
        \bigg]
        \\ 
        &\quad\quad\quad\times
        \exp N \log
        \Bigg[
            \sum_{\ux} 
            e^{               
                - i \nu_\epsilon \sum_{t < s} \hQ(t,s) x^t x^s
            }
            \int d\uh \,
            \theta_{\caT, \sigma}(\ux, \uh) e^{
                g(\ux, \uh)
            }
            \\&\quad\quad\quad\quad\times
            \caN\bigg( \uh ; - i \nu_\epsilon  \sum_{w} \hS(w,s) x^w ; 
                    - i \nu_\epsilon \big(
                    \delta(t=s) 
                    2 \hR(t,t)
                    + \delta(t\neq s) \hR(t,s) 
                \big)
            \bigg)
        \Bigg]
    \end{split}
\end{equation}

Thus, we arrived at a first saddle-point representation for the annealed entropy as a function of the order parameters $Q,R,S$ and their conjugate variables $\hQ, \hR, \hS$:
\begin{equation}
    \begin{split}
        s_{\rm ann}(\caT, g)
        &\approx
        \max_{\text{op.}}
        \Bigg\{
        \nu_\epsilon
        \bigg[
            \sum_{t < s} Q(t,s) R(t,s)
            + \frac{1}{2} \sum_{t} R(t,t)
            + \frac{1}{2} \eta_\epsilon \sum_{t,s} S(t,s) S(s,t)
        \bigg]
        \\ 
        &\quad\quad\quad+
        \nu_\epsilon 
        \bigg[
            i \sum_{t < s}\hQ(t,s) Q(t,s) 
            + i \sum_{t \leq s}\hR(t,s) R(t,s) 
            + i \sum_{t ,s}\hS(t,s) S(t,s) 
        \bigg]
        \\ 
        &\quad\quad\quad+
        \log
        \Bigg[
            \sum_{\ux} 
            e^{               
                - i \nu_\epsilon \sum_{t < s} \hQ(t,s) x^t x^s
            }
            \int d\uh \,
            \theta_{\caT, \sigma}(\ux, \uh) e^{
                g(\ux, \uh)
            }
            \\&\quad\quad\quad\quad\times
            \caN\bigg( \uh ; - i \nu_\epsilon  \sum_{w} \hS(w,s) x^w ; 
                    - i \nu_\epsilon \big(
                    \delta(t=s) 
                    2 \hR(t,t)
                    + \delta(t\neq s) \hR(t,s) 
                \big)
            \bigg)
        \Bigg]
        \Bigg\}
    \end{split}
\end{equation}

\subsection{Saddle point equations I}

We now notice that the annealed entropy depends only quadratically on the non-hat order parameters.
Thus, we can start performing the saddle-point maximisation on the non-hat variables.
The stationarity conditions imply
\begin{equation}
    \begin{split}
        \del_{Q(t,s)} &= 0
        \implies 
        \nu_\epsilon R(t,s) + i \nu_\epsilon \hQ(t,s) = 0
        \implies
        i \hQ(t,s) = -R(t,s) 
        \mathfor t<s
        \, ,
        \\
        \del_{R(t,s)} &= 0
        \implies 
        \nu_\epsilon Q(t,s) + i \nu_\epsilon \hR(t,s) = 0
        \implies
        i\hR(t,s) = - Q(t,s) 
        \mathfor t<s
        \, ,
        \\
        \del_{R(t,s)} &= 0
        \implies 
        \nu_\epsilon / 2 + i \nu_\epsilon \hR(t,t) = 0
        \implies
        i\hR(t,t) = - 1/2
        \mathfor t=s
        \, ,
        \\
        \del_{S(t,s)} &= 0
        \implies 
        \nu_\epsilon \eta_\epsilon S(s,t) + i \nu_\epsilon \hS(t,s) = 0
        \implies
        i \hS(t,s) = - \eta_\epsilon S(s,t)
        \mathfor t,s
        \, ,
    \end{split}
\end{equation}
These equations give us explicit values for the non-hat order parameters. They also provide a bijection between hat and non-hat order parameters. Thus, we can directly solve for the hat variables, obtaining
\begin{equation}
    \begin{split}
        s_{\rm ann}(\caT, g)
        &\approx
        \max_{Q,R,S}
        \Bigg\{
        \nu_\epsilon
        \bigg[
            - \sum_{t < s} Q(t,s) R(t,s)
            - \frac{1}{2} \eta_\epsilon \sum_{t,s} S(t,s) S(s,t)
        \bigg]
        \\ 
        &\quad\quad\quad+
        \log
        \Bigg[
            \sum_{\ux} 
            e^{               
                \nu_\epsilon \sum_{t < s} R(t,s) x^t x^s
            }
            \int d\uh \,
            \theta_{\caT, \sigma}(\ux, \uh) e^{
                g(\ux, \uh)
            }
            \caN\bigg( \uh ; \eta_\epsilon \nu_\epsilon S \ux ; \nu_\epsilon (\mathbb{I} + Q) 
            \bigg)
        \Bigg]
        \Bigg\} \, ,
    \end{split}
\end{equation}
where $\mathbb{I} + Q$ represents the symmetric matrix with unit diagonal and matching the order parameter $Q$ out of diagonal (recall that $Q(t,s)$ is defined only for $t<s$).
Also, notice that the dependence on $R(t,t)$ dropped out of the exponent at leading order, meaning that $R(t,s)$ should be now thought as defined only for $t<s$.
Now, it can be useful to map 
\begin{equation}\label{op.ch1}
    \begin{split}
        \uh \to \nu_\epsilon^{1/2} \uh \, ,\quad 
        R \to \nu_\epsilon^{-1} R \mathand
        S \to \nu_\epsilon^{-1/2} S \, ,
    \end{split}
\end{equation}
so that (we are discaring here constant factors coming from the change of variable as they are subleading)
\begin{equation}
    \begin{split}
        s_{\rm ann}(\caT, g)
        &\approx
        \max_{Q,R,S}
        \Bigg\{
            - \sum_{t < s} Q(t,s) R(t,s)
            - \frac{\eta_\epsilon}{2} \sum_{t,s} S(t,s) S(s,t)
        \\ 
        &\quad\quad\quad+ \log
        \Bigg[
            \sum_{\ux} 
            e^{\sum_{t < s} R(t,s) x^t x^s}
            \int d\uh \,
            \theta_{\caT, \sigma}(\ux, \sqrt{\nu_\epsilon} \, \uh) e^{
                g(\ux, \sqrt{\nu_\epsilon} \, \uh)
            }
            \caN\bigg( 
                \sqrt{\nu_\epsilon} \uh ; 
                \sqrt{\nu_\epsilon} 
                \eta_\epsilon S \ux ; 
                \nu_\epsilon (\mathbb{I} + Q) 
            \bigg)
        \Bigg]
        \Bigg\}
        \\
        &\approx
        \max_{Q,R,S}
        \Bigg\{
            - \sum_{t < s} Q(t,s) R(t,s)
            - \frac{\eta_\epsilon}{2} \sum_{t,s} S(t,s) S(s,t)
        \\ 
        &\quad\quad\quad+ \log
        \bigg[
            \sum_{\ux} 
            e^{\sum_{t < s} R(t,s) x^t x^s}
            \int d\uh \,
            \theta_{\caT, \sigma}(\ux, \sqrt{\nu_\epsilon} \, \uh) e^{
                g(\ux, \sqrt{\nu_\epsilon} \, \uh)
            }
            \caN\big( 
                \uh ; 
                \eta_\epsilon S \ux ; 
                \mathbb{I} + Q
            \big)
        \bigg]
        \Bigg\} \, .
    \end{split}
\end{equation}
Finally, we notice that if $\eta_\epsilon = 0$, then the dependence on $S$ drops. We can alternatively assume that $S= 0$ is the SP, as any value for $S$ is a SP in that case.
If $\eta_\epsilon \neq 0$, we perform the change of variable 
\begin{equation} \label{op.ch2}
    \eta_\epsilon S \to V \, ,
\end{equation}
so that (we are discarding here constant factors in front of the integrals as they are subleading)
\begin{equation}
    \begin{split}
        s_{\rm ann}(\caT, g)
        &\approx
        \max_{Q,R,V}
        \Bigg\{
            - \sum_{t < s} Q(t,s) R(t,s)
            - \frac{1}{2\eta_\epsilon} \sum_{t,s} V(t,s) V(s,t)
        \\ 
        &\quad\quad\quad+ \log
        \bigg[
            \sum_{\ux} 
            e^{\sum_{t < s} R(t,s) x^t x^s}
            \int d\uh \,
            \theta_{\caT, \sigma}(\ux, \sqrt{\nu_\epsilon} \, \uh) e^{
                g(\ux, \sqrt{\nu_\epsilon} \, \uh)
            }
            \caN\big( 
                \uh ; 
                V \ux ; 
                \mathbb{I} + Q
            \big)
        \bigg]
        \Bigg\}
    \end{split}
\end{equation}

We can write this more compactly by considering $Q$ and $R$ as $T \times T$ symmetric matrices with zero diagonal, giving
\begin{equation}
    \begin{split}
        s_{\rm ann}(\caT, g)
        &\approx
        \max_{Q,R,V}
        \Bigg\{
            - \frac{1}{2} \Tr\left( Q R + \frac{1}{\eta_\epsilon} V V^T \right)
        % \\ 
        % &\quad\quad\quad
        + \log
        \bigg[
            \sum_{\ux} 
            e^{\frac{1}{2} \ux^T R \ux }
            \int d\uh \,
            \theta_{\caT, \sigma}(\ux, \sqrt{\nu_\epsilon} \, \uh) e^{
                g(\ux, \sqrt{\nu_\epsilon} \, \uh)
            }
            \caN\big( 
                \uh ; 
                V \ux ; 
                \mathbb{I} + Q
            \big)
        \bigg]
        \Bigg\} \, .
    \end{split}
\end{equation}

Notice that the logarithmic part of the annealed entropy defines a stochastic process $(\ux, \uh) \in \{\pm 1\}^{T+1} \times \bbR^T$ with probability density (here we restrict to $g=0$, as this will be the case whenever considering the stochastic process)
\begin{equation}\label{eq.stoc}
    p(\ux, \uh) = \frac{1}{\caZ} e^{\frac{1}{2} \ux^T R \ux } \,
    \theta_{\caT, \sigma}(\ux, \sqrt{\nu_\epsilon} \, \uh) \,
    \caN\big( 
        \uh ; 
        V \ux ; 
        \mathbb{I} + Q
    \big)
\end{equation}
with associated average
\begin{equation}
    \angavg{h(\ux, \uh)} = \sum_{\ux} \int d\uh \, h(\ux, \uh) \, p(\ux, \uh) \, ,
\end{equation}
and normalisation factor
\begin{equation}
    \caZ = \sum_{\ux} \int d\uh \,
    e^{\frac{1}{2} \ux^T R \ux } \,
    \theta_{\caT, \sigma}(\ux, \sqrt{\nu_\epsilon} \, \uh) \,
    \caN\big( 
        \uh ; 
        V \ux ; 
        \mathbb{I} + Q
    \big)
\end{equation}

Recalling that 
\begin{equation}
    g(\ux_i, \uh_i) = 
    \sum_{t < s} \omega^{xx}_{t,s} x_i(t) x_i(s)
    + \sum_{t,s} \omega^{xh}_{t,s} x_i(t) h_i(s)
    + \sum_{t \leq s} \omega^{hh}_{t,s} h_i(t) h_i(s)
\end{equation}
the observables we are interested in, namely the dynamical correlators and the energy, can be computed (after completing the maximisation over $Q,R,V$) as
\begin{equation}
    \begin{split}
        C^{xx}(t,s) &= \angavg{x(t)x(s)}
        \, ,\\
        C^{xh}(t,s) &= \sqrt{\nu_\epsilon} \angavg{x(t)h(s)}
        \, ,\\
        C^{hh}(t,s) &= \nu_\epsilon \angavg{h(t)h(s)}
        \, , 
    \end{split}
\end{equation}
and
\begin{equation}
    E(t) = - \frac{1}{2} C^{xh}(t,t) = - \frac{\sqrt{\nu_\epsilon}}{2} \angavg{x(t)h(t)} \, ,
\end{equation}
(each of the angular averages is performed at $g=0$).

Thus, the stochastic process can be thought of as a mean-field representative of the dynamics of the spins and of the local fields. 
As we could have expected, the local fields are Gaussians, and the spin-to-spin coupling is replaced in mean-field by a time-to-time correlation mediated by the order parameters $Q, R, V$.

\subsection{Saddle point equations II}

We now derive the saddle-point equations for $Q,R,V$. 
We fix $g=0$ from now on, as we are not interested in tuning the values of the dynamical correlators.

The saddle-point equation obtained taking the derivative w.r.t. $R(t,s)$ is simple and gives, for $t<s$, 
\begin{equation}
    Q(t,s) = \angavg{x^t x^s} \, .
\end{equation}
We provide two alternative representations for the saddle-point equations obtained taking the derivative w.r.t. $Q, V$.

\subsubsection{First representation}
For the first representation, it's best to work in the Fourier representation
\begin{equation}
    \begin{split}
        \caN\big( 
            \uh ; 
            V \ux ; 
            \mathbb{I} + Q
        \big) 
        &= 
        \int \frac{d\ulambda}{(2\pi)^{T}} \, 
        e^{
            -\frac{1}{2} \ulambda^T (\mathbb{I} + Q) \ulambda 
            + i \ulambda^T \uh
            - i \ulambda^T V \ux
            }
    \end{split}
\end{equation}
so that the equation for $\del_{V(t,s)}$ gives (call $\caZ$ the normalisation of the angular average)
\begin{equation}
    \begin{split}
        V(s,t) &=
        \eta_\epsilon \caZ^{-1}
        \sum_{\ux} 
        e^{\frac{1}{2} \ux^T R \ux}
        \int \frac{d\uh \,d\ulambda}{(2\pi)^{T}} 
        \theta_{\caT, \sigma}(\ux, \sqrt{\nu_\epsilon} \, \uh) 
        \del_{V(t,s)}
        e^{
            -\frac{1}{2} \ulambda^T (\mathbb{I} + Q) \ulambda 
            + i \ulambda^T \uh
            - i \ulambda^T V \ux
        }
        \\
        &=
        - \eta_\epsilon \caZ^{-1}
        \sum_{\ux} 
        x_s
        e^{\frac{1}{2} \ux^T R \ux}
        \int \frac{d\uh \,d\ulambda}{(2\pi)^{T}} 
        \theta_{\caT, \sigma}(\ux, \sqrt{\nu_\epsilon} \, \uh) 
        (i \lambda_t) 
        e^{
            -\frac{1}{2} \ulambda^T (\mathbb{I} + Q) \ulambda 
            + i \ulambda^T \uh
            - i \ulambda^T V \ux
        }
        \\
        &=
        - \eta_\epsilon \caZ^{-1}
        \sum_{\ux} 
        x_s
        e^{\frac{1}{2} \ux^T R \ux}
        \int \frac{d\uh \,d\ulambda}{(2\pi)^{T}} 
        \theta_{\caT, \sigma}(\ux, \sqrt{\nu_\epsilon} \, \uh) 
        \del_{h_t}
        e^{
            -\frac{1}{2} \ulambda^T (\mathbb{I} + Q) \ulambda 
            + i \ulambda^T \uh
            - i \ulambda^T V \ux
        }
        \\
        &=
        - \eta_\epsilon \caZ^{-1}
        \sum_{\ux} 
        x_s
        e^{\frac{1}{2} \ux^T R \ux}
        \int d\uh 
        \theta_{\caT, \sigma}(\ux, \sqrt{\nu_\epsilon} \, \uh) 
        \del_{h_t}
        \caN\big( 
            \uh ; 
            V \ux ; 
            \mathbb{I} + Q
        \big)
        \\
        &=
        \eta_\epsilon \caZ^{-1}
        \sum_{\ux} 
        x_s
        e^{\frac{1}{2} \ux^T R \ux}
        \int d\uh \,
        \caN\big( 
            \uh ; 
            V \ux ; 
            \mathbb{I} + Q
        \big) \,
        \del_{h_t}
        \theta_{\caT, \sigma}(\ux, \sqrt{\nu_\epsilon} \, \uh) 
         \, .
    \end{split}
\end{equation}
The equation 
for $\del_{Q(t,s)}$ with $t<s$ gives 
\begin{equation}
    \begin{split}
        R(t,s) &=
        \caZ^{-1}
        \sum_{\ux} 
        e^{\frac{1}{2} \ux^T R \ux}
        \int \frac{d\uh \,d\ulambda}{(2\pi)^{T}} 
        \theta_{\caT, \sigma}(\ux, \sqrt{\nu_\epsilon} \, \uh) 
        \del_{Q(t,s)}
        e^{
            -\frac{1}{2} \ulambda^T (\mathbb{I} + Q) \ulambda 
            + i \ulambda^T \uh
            - i \ulambda^T V \ux
        }
        \\
        &=
        \caZ^{-1}
        \sum_{\ux} 
        e^{\frac{1}{2} \ux^T R \ux}
        \int \frac{d\uh \,d\ulambda}{(2\pi)^{T}} 
        \theta_{\caT, \sigma}(\ux, \sqrt{\nu_\epsilon} \, \uh) 
        (- \lambda_t \lambda_s) 
        e^{
            -\frac{1}{2} \ulambda^T (\mathbb{I} + Q) \ulambda 
            + i \ulambda^T \uh
            - i \ulambda^T V \ux
        }
        \\
        &=
        \caZ^{-1}
        \sum_{\ux} 
        e^{\frac{1}{2} \ux^T R \ux}
        \int \frac{d\uh \,d\ulambda}{(2\pi)^{T}} 
        \theta_{\caT, \sigma}(\ux, \sqrt{\nu_\epsilon} \, \uh) 
        \del_{h_t, h_s}
        e^{
            -\frac{1}{2} \ulambda^T (\mathbb{I} + Q) \ulambda 
            + i \ulambda^T \uh
            - i \ulambda^T V \ux
        }
        \\
        &=
        \caZ^{-1}
        \sum_{\ux} 
        e^{\frac{1}{2} \ux^T R \ux}
        \int d\uh 
        \theta_{\caT, \sigma}(\ux, \sqrt{\nu_\epsilon} \, \uh) 
        \del_{h_t, h_s}
        \caN\big( 
            \uh ; 
            V \ux ; 
            \mathbb{I} + Q
        \big)
        \\
        &=
        \caZ^{-1}
        \sum_{\ux} 
        e^{\frac{1}{2} \ux^T R \ux}
        \int d\uh \,
        \caN\big( 
            \uh ; 
            V \ux ; 
            \mathbb{I} + Q
        \big) \,
        \del_{h_t, h_s}
        \theta_{\caT, \sigma}(\ux, \sqrt{\nu_\epsilon} \, \uh) \, .
    \end{split}
\end{equation}

\subsubsection{Second representation}
For the second representation we use
\begin{equation}
    \begin{split}
        \del_{\phi_i} \caN(\phi; 0, \Sigma) 
        &=
        (\Sigma^{-1} \phi)_i \caN(\phi; 0, \Sigma) \\
        \del_{\mu_i} \caN(\phi; 0, \Sigma) 
        &= 
        (\Sigma^{-1} \phi)_i \caN(\phi; 0, \Sigma) \\
        \del_{\Sigma_{ij}} \caN(\phi; 0, \Sigma) 
        &= 
        \left(\frac{1}{2} (\Sigma^{-1} \phi)_i ((\Sigma^{-1} \phi))_j 
        -\frac{1}{2} \Sigma^{-1}_{ij} \right)
        \caN(\phi; 0, \Sigma) 
    \end{split}
\end{equation}
where we call $\uphi = \uh - V\ux$.

For the SP equation for $V(t,s)$ for all $t,s$, we obtain
\begin{equation}\label{eq.Vxh}
    \begin{split}
        V(s,t) &= \eta_\epsilon
        \angavg{
            \sum_{u,v} 
            (\mathbb{I}+Q)^{-1}_{uv} \phi_v
            \del_{V(t,s)} \sum_{y} V(u,y) x_y
        }
        \\
        &= \eta_\epsilon
        \angavg{
            \sum_{u,v} 
            (\mathbb{I}+Q)^{-1}_{uv} 
            \phi_v
            \sum_{y} \delta_{tu} \delta_{sy} x_y
        }
        \\
        &= \eta_\epsilon    
            \sum_{v} 
            \angavg{x_s \phi_v}
            (\mathbb{I}+Q)^{-1}_{vt} 
        \\      
        &\implies 
        \angavg{x_t \phi_s} = \eta_\epsilon^{-1} \sum_v V(t, v) (\mathbb{I}+Q)_{v,s}
    \end{split}
\end{equation}
This also provides an expression for the energy as a function of the order parameters
\begin{equation}
    \begin{split}
        E(t) 
        &= C^{xh}(t,t) 
        = \frac{\sqrt{\nu_\epsilon}}{2} \angavg{x_t h_t }
        = \frac{\sqrt{\nu_\epsilon}}{2} \angavg{x_t  (\phi_t  + (V\ux)_t) }
        \\
        &=
        \frac{\sqrt{\nu_\epsilon}}{2}
        \left[
            \angavg{x_t\phi_t} + \sum_{v} V(t,v) \angavg{x_t x_v}
        \right]
        \\
        &=
        \frac{\sqrt{\nu_\epsilon}}{2}
        \left[
            \eta_\epsilon^{-1} \sum_v V(t, v) (\mathbb{I}+Q)_{v,t} + \sum_{v} V(t,v) (\mathbb{I}+Q)_{v, t}
        \right]
        \\
        &=
        \frac{\sqrt{\nu_\epsilon}}{2} \,
        \frac{1+\eta_\epsilon}{\eta_\epsilon}  \, 
        \sum_{v} 
        (\mathbb{I}+Q)_{tv}V(t,v)
    \end{split}
\end{equation}

For the SP for $Q(t,s)$ for $t<s$ we obtain
\begin{equation}
    \begin{split}
        R(t,s) &=
        \frac{1}{2} 
        \sum_{uv}
        \angavg{
            \left(
            ((\mathbb{I}+Q)^{-1} \phi)_u
            ((\mathbb{I}+Q)^{-1} \phi)_v 
            - (\mathbb{I}+Q)^{-1}_{uv} 
            \right)
            \del_{Q(t,s)} (\mathbb{I}+Q)_{uv}
        }
        \\&=
        \sum_{u<v}
        \angavg{
            \left(
            ((\mathbb{I}+Q)^{-1} \phi)_u
            ((\mathbb{I}+Q)^{-1} \phi)_v 
            - (\mathbb{I}+Q)^{-1}_{uv} 
            \right)
            \del_{Q(t,s)} (\mathbb{I}+Q)_{uv}
        }
        \\&=
        \sum_{u<v}
        \angavg{
            \left(
            ((\mathbb{I}+Q)^{-1} \phi)_u
            ((\mathbb{I}+Q)^{-1} \phi)_v 
            - (\mathbb{I}+Q)^{-1}_{uv} 
            \right)
            \delta_{tu} \delta_{sv}
        }
        \\&=
        \frac{1}{2}
        \sum_{u \neq v}
        \angavg{
            \left(
            ((\mathbb{I}+Q)^{-1} \phi)_u
            ((\mathbb{I}+Q)^{-1} \phi)_v 
            - (\mathbb{I}+Q)^{-1}_{uv} 
            \right)
            (\delta_{tu} \delta_{sv} + \delta_{su} \delta_{tv})
        }
        \\&=
        \sum_{ab}
        (\mathbb{I}+Q)^{-1}_{ta}
        \angavg{\phi_a\phi_b}
        (\mathbb{I}+Q)^{-1}_{bs}
        - (\mathbb{I}+Q)^{-1}_{ts} 
        \\&\implies
        \angavg{\phi_t\phi_s} 
        = \left[(\mathbb{I}+Q)(R + (\mathbb{I}+Q)^{-1} )(\mathbb{I}+Q)\right]_{ts}
        = \left[(\mathbb{I}+Q)R(\mathbb{I}+Q) + \mathbb{I}+Q\right]_{ts}
    \end{split}
\end{equation}
where in the second passage we used that $\del_Q (\mathbb{I}+Q)_{uu} = 0$ + symmetry.

Thus, we see that $V$ and $R$ can be found by computing shifted correlators $\angavg{x \phi}$ and $\angavg{\phi\phi}$, and inverting a linear system.
This representation also gives directly a relationship between the order parameters and the dynamical correlators.

\subsection{Simplifications in DMFT for forward dynamics}
\label{app.fd}

For the case of forward dynamics, $\caT = \caT_{\rm fd} = {\pm 1}^{T+1}$, a number of simplifications arise.
We consider the following ansatz (which we call "causal ansatz") for the order parameters, and verify that it satisfies the SP equations: 
$R(t,s) = 0$ for all $(t,s)$, and $V(t,s) = 0$ for all $t \leq s$.
Notice that this can be seen as a consequence of causality, i.e. that perturbations in the fields at larger times cannot have effect on the spins at previous times, and that perturbations in the fields at different times are uncorrelated.

The saddle-point equations under the causal ansatz are
\begin{equation}
    \begin{split}
        \caZ &= \sum_{\ux} \int d\uh \,
        \caN\left( 
            \uh ; 
            \left( \sum_{v<u} V(u,v) x_v \right)_u ; 
            \mathbb{I} + Q
        \right) \times
        \theta_{\caT_{\rm fd}, \sigma}(\ux, \sqrt{\nu_\epsilon} \, \uh) 
        \\
        Q(t,s) &= \caZ^{-1} \sum_{\ux} \int d\uh \,
        \caN\left( 
            \uh ; 
            \left( \sum_{v<u} V(u,v) x_v \right)_u ; 
            \mathbb{I} + Q
        \right) \times
        x_s x_t
        \theta_{\caT_{\rm fd}, \sigma}(\ux, \sqrt{\nu_\epsilon} \, \uh) 
        \\
        V(s,t) &= \eta_\epsilon \caZ^{-1} \sum_{\ux} \int d\uh \,
        \caN\left( 
            \uh ; 
            \left( \sum_{v<u} V(u,v) x_v \right)_u ; 
            \mathbb{I} + Q
        \right) \times
        x_s \del_{h_t}
        \theta_{\caT_{\rm fd}, \sigma}(\ux, \sqrt{\nu_\epsilon} \, \uh) 
        \\
        R(t,s) &= \caZ^{-1} \sum_{\ux} \int d\uh \,
        \caN\left( 
            \uh ; 
            \left( \sum_{v<u} V(u,v) x_v \right)_u ; 
            \mathbb{I} + Q
        \right) \times
        \del_{h_s} \del_{h_t}
        \theta_{\caT_{\rm fd}, \sigma}(\ux, \sqrt{\nu_\epsilon} \, \uh) 
    \end{split}
\end{equation}
and recall that 
\begin{equation}
    \theta_{\caT_{\rm fd}, \sigma}(\ux,  \sqrt{\nu_\epsilon} \,\uh) 
    =
    \prod_{t=1}^{T} 
    \delta\bigg[x_{t+1} = \sigma\Big( x_t,  \sqrt{\nu_\epsilon} \,h_t \Big) \bigg] \, .
\end{equation}

Under the causal ansatz, it is immediate to see that $\caZ = 2$.
Indeed, one can sum over $x_{T+1}$, which appears only in the dynamical constraint, giving
\begin{equation}
    \sum_{x_{T+1}} \delta( x_{T+1} = \sigma(x_T, h_T)) = 1 \, .
\end{equation}
Then, the integration over $h_T$ (which is not anymore constrained thanks to the summation over $x_{T+1}$) can be performed: one simply drops the $T$-th dimension in the multivariate Gaussian, thanks to its marginalisation properties. 
Due to the triangularity of $V$, we see that dropping the $T$-th dimension makes the Gaussian measure independent on $x_T$.
This leads to a $\caZ$ structurally identical to the original one, but with one less time step. Iterating the summation over $x_{t+1}$ and the integration over $h_t$ for $t=T, T-1, \dots, 1$ gives
\begin{equation}
    \caZ = \sum_{x_1} 1 = 2 \, .
\end{equation}
Notice that the summation over $x_1$ could have been performed since the start, as the measure $p(\ux, \uh)$ is invariant under inversion $p(\ux, \uh) = p(-\ux, -\uh)$.
This, together with the causal ansatz, readily implies that
\begin{equation}
    s_{\rm ann}(\caT_{\rm fd}) = \log 2 \, ,
\end{equation}
as expected.

It is now useful to look explicitly at
\begin{equation}\label{eq.fd-constr-del}
    \begin{split}
        \del_{h_t} \theta_{\caT_{\rm fd}, \sigma}(\ux,  \sqrt{\nu_\epsilon} \,\uh) 
        &=
        \del_{h_t}\delta\bigg[x_{t+1} = \sigma\Big( x_t,  \sqrt{\nu_\epsilon} \,h_t \Big) \bigg]
        \prod_{u=1, u \neq t}^{T} 
        \delta\bigg[x_{u+1} = \sigma\Big( x_u,  \sqrt{\nu_\epsilon} \,h_u \Big) \bigg] 
        \\
        &=
        x_{t+1} \left(\sum_{z \in I_+(x_t)} \delta(h_t - z /  \sqrt{\nu_\epsilon} ) - \sum_{z \in I_-(x_t)} \delta(h_t - z /  \sqrt{\nu_\epsilon} ) \right)
        \prod_{u=1, u \neq t}^{T} 
        \delta\bigg[x_{u+1} = \sigma\Big( x_u, h_u \Big) \bigg] 
        \, .
    \end{split}
\end{equation}
where the sum over $z$ runs over $I_+(x_t)$, the set of points where the function $h_t \to \sigma(x_t, h_t)$ changes sign from $-1$ to $+1$, and over $I_-(x_t)$, the set of points where the function $h_t \to \sigma(x_t, h_t)$ changes sign from $+1$ to $-1$.
The sum over the deltas is not important at this stage, the important part is that the factor 
\begin{equation}
    x_{t+1}  \delta\bigg[x_{t+2} = \sigma\Big( x_{t+1}, h_{t+1} \Big) \bigg]
\end{equation}
contains the only dependency on $x_{t+1}$ in the whole constraint function.

Let us look at the equation for $R(t,s)$ for $t<s$. 
As for the normalisation factor $\caZ$, we can immediately sum over $x_{u+1}$ and integrate over $h_u$ for all $u = s+2, \dots, T+1$.  
Then, when summing over $x_{s+1}$, we have 
\begin{equation}
    \sum_{x_{s+1}} x_{s+1} = 0 \, ,
\end{equation}
giving immediately $R(t,s) = 0$.

For $V(s,t)$, the same reasoning as for $R$ applies as long as $t \geq s$, as in that case the derivative $\del_{h_t}$ produces a term $x_{t+1}$ that behaves as above for $R$.

Thus, we have shown that the causal ansatz is indeed consistent with the saddle-point equations. 

It is easy to see that the equation for $Q$ and for the non-trivial triangle of $V$ are non-trivial, as no clear simplification arises when summing 
\begin{equation}
    \sum_{x_{t}} x_{t} \delta\bigg[x_{t} = \sigma\Big( x_{t-1}, h_{t-1} \Big) \bigg] = \, ??? \, .
\end{equation}
Nonetheless, we see that in both equations for $Q(t,s)$ and $V(s,t)$ with $s>t$ we can sum over all times $u = s+1, \dots, T+1$, so that, for example, $Q(t,s)$ depends only on $\{Q(u,v), V(v,u) \mathfor 1\leq v<u<s \}$.
Thus, the saddle-point equations can be solved directly in a forward scheme, in which $\{Q(1,2), V(2,1)\}$ are computed, then $\{ Q(1,3), Q(2,3), V(3,1), V(3,2) \}$ are computed as a function of $\{Q(1,2), V(2,1)\}$, etc \dots.

In the most simplified form, the equations for $Q(t,s)$ and $V(s,t)$ read, for $s>t$
\begin{equation}\label{eq.fd-dmft}
    \begin{split}
        Q(t,s) &= \sum_{\ux} \int d\uh \,
        \caN\left( 
            \uh ; 
            \left( \sum_{v<u} V(u,v) x_v \right)_u ; 
            \mathbb{I} + Q
        \right) \times
        x_t
        \theta_{\caT_{\rm fd}, \sigma}(\ux, \sqrt{\nu_\epsilon} \, \uh) 
        \\
        V(s,t) &= \eta_\epsilon \sum_{\ux} \int d\uh \,
        \caN\left( 
            \uh ; 
            \left( \sum_{v<u} V(u,v) x_v \right)_u ; 
            \mathbb{I} + Q
        \right) \times
        \del_{h_t}
        \theta_{\caT_{\rm fd}, \sigma}(\ux, \sqrt{\nu_\epsilon} \, \uh) 
        \\
    \end{split}
\end{equation}
where the integral is over $\uh \in \bbR^{s-1}$, the sum is over $\ux \in \{\pm 1\}^{s-1}$ and we used the inversion symmetry of $p(\ux, \uh)$ to fix $x_s = 1$ simplifying away the normalisation $\caZ$ (but we could have fixed any other spin to 1).
For $t=1, s=2$ we have
\begin{equation}
    \begin{split}
        Q(1,2) &= \sum_{x=\pm 1} x \int dh \,
        \caN\left( h ; 0 ; 1\right)
        \delta(\sigma(x, h)=1)
        \\
        V(2,1) &= \eta_\epsilon \sum_{x=\pm 1} \int dh \,
        \caN\left( h ; 0 ; 1\right)
        \del_h
        \delta(\sigma(x, h)=1)
        \\
    \end{split}
\end{equation}
which are easily computable in terms of the p.d.f. and c.d.f. of the standard Gaussian functions for all dynamical rules $\sigma$ which are well-behaved (meaning for example that $h \to \sigma(x,h)$ changes sign on a zero-measure set for both $x=\pm 1$). 

Moreover, we notice that sampling the stochastic process in the forward dynamics case is easy. 
Indeed, one can sample from $p(\ux, \uh)$ for given $Q$ and $V$ by:
\begin{itemize}
    \item either sampling $\uphi = \uh - V\ux$ from its multivariate Gaussian (independent on $\ux$), then setting $x_1 = \pm 1$ with equal probabilities and running the dynamical rule $x_{t+1} = \sigma(x_t, \phi_t + (V\ux)_t)$ for $t=1, \dots, T$, where we stress that $(V\ux)_t$ depends only on $(x_1, \dots, x_{t-1})$ by causality of $V$.
    \item Or sampling $x_1 = \pm 1$ with equal probability, then sampling $\phi_1 = h_1 - (V\ux)_1$ from the corresponding marginal of the multivariate Gaussian, then computing $x_2 = \sigma(x_1, \phi_1 + (V\ux)_1)$, then sampling $\phi_2 = h_2 - (V\ux)_2$ given $\phi_1$, and so on up to $\phi_T$ and $x_{T+1}$.
        To this end, it is useful to know that the multivariate Gaussian is well-behaved under conditioning. In fact
            \begin{equation}
                \caN(x) = \caN_{1|2}(x_1 | x_2) \caN_*(x_2)
            \end{equation}
            where $\caN_*$ is the marginal over the subset of variables $x_2$ (a Gaussian) and $\caN_{1|2}$ is the conditional probability of the remaining variables $x_1$ after conditioning over $x_2$ (again, non-trivially, a Gaussian).
            In paricular, we have, decomposing the mean vector and covariance matrix of $\caN$ as
            \begin{equation}
                \mu = \begin{bmatrix}
                    \mu_1 \\ \mu_2
                \end{bmatrix}
                \mathand
                \Sigma = \begin{bmatrix}
                    \Sigma_{11} & \Sigma_{12} \\
                    \Sigma_{12}^T & \Sigma_{22}
                \end{bmatrix}
                \, ,
            \end{equation}
            we have that the mean and covariance of the measures $\caN_*$ and $\caN_{1|2}$ satisfy
            \begin{equation}
                \begin{split}
                    \mu_* &= \mu_2 \, ,
                    \\
                    \Sigma_* &= \Sigma_{22} \, ,
                    \\
                    \mu_{1|2}
                    &= \mu_1 + \Sigma_{12} \Sigma_{22}^{-1} (x_2 - \mu_2) \, ,
                    \\
                    \Sigma_{1|2}
                    &= \Sigma_{11} - \Sigma_{12} \Sigma_{22}^{-1} \Sigma_{12}^T \, .
                \end{split}
            \end{equation}
            This allows to perform the conditional sampling procedure described above efficiently.
\end{itemize}

As a final remark, we recall that for the sync-Glauber dynamics $\sigma(x,h) = \sign(h)$ the order parameters $Q$ and $V$ satisfy a peculiar symmetry \cite{opper94}, that states that $Q$ and $V$ are alternatively null, implying that the energy $E(t) = 0$ for all times $t$.

\subsection{Complications in DMFT for backtracking dynamics}

\subsubsection{Lack of simplifications analogous to the forward DMFT}

All the simplifications discussed above do not hold in the case of $(p,c)$-backtracking dynamics.
In particular
\begin{itemize}
    \item the order parameters are non-causal, i.e. $R(t,s) \neq 0$ for $t<s$ and $V(t,s) \neq 0$ for $t\leq s$. This is due to the cyclicity constraint $x(p+1) = x(T+1)$ with $T=p+c$, which prevents the summation/integration trick seen in the previous section.
    \item The normalisation factor $\caZ$ is non-trivial, due to the non-causality of the order parameters.
    \item The stochastic process is difficult to sample. Referring with the two sampling strategies discussed in the previous section, the main difficulty to overcome is the $V \ux$ term in the mean of the Gaussian: this is now non-causal, meaning that $h_t$ for all $t$ depends on the full spin trajectory $\ux$. This prevents from just running the dynamics to satisfy the compatibility constraint between $\uh$ and $\ux$.
    \item the saddle-point equations are not explicit as in the forward case, meaning that they need to be solved numerically (usually by a fixed-point iteration scheme, as usual in replica-style computations).
\end{itemize}
All these considerations make the backtracking DMFT severely more challenging to analyse, and the relevant equations more difficult to solve numerically.

Notice that in the $p=0$, $c\geq 1$ case considered in \cite{hwang2019number} for the sync-Glauber dynamics $\sigma(x,h) = \sign(h)$, additional symmetries allow to somewhat simplify the backtracking saddle-point equations. This is not the case as soon as $p\geq 1$, or for more complicated dynamical rules.

\subsubsection{Degeneracy}\label{app.degeneracy}

We point out the fact that there is some degeneracy between the $(p,c)$-backtracking entropy, and another $(p', c')$-backtracking entropy in general.
This has been discussed in detail for the case $c=0$ in \cite{hwang2019number}, to which we refer the interested reader. 
Here we add some specific detail for the case $c=1,2$.

We observed that each saddle-point at a given $p$ and $c=1$ induces automatically a saddle-point for $p'=p$ and $c'=2$. 
This can be seen intuitively by arguing that $(p,2)$-backtracking attractors include as a subset $(p,1)$-backtracking attractors, as a 1-cycle is also a trivial 2-cycle. What is less trivial is the fact that $(p,1)$ attractors and $(p,2)$ attractors lead to two distinct solutions of the $(p,2)$ saddle point equations, one describing the trivial 2-cycles, and one describing the non-trivial 2-cycles.

More technically, we can consider the stochastic process \eqref{eq.stoc}, and look at the ansatz $Q(T-1, T)=1$ and $Q(t,T-1) = Q(t, T)$ for all $1 \leq t < T-2$, where $T = p'+c' = p+2$. We can verify that this ansatz is consistent and leads to a saddle-point equations system identical to the one for $(p, c=1)$. 
The ansatz implies that the Gaussian measure for $h_T$ conditioned on $\uh_{<T}$ and $\ux$ is singular, and equals $\delta(h_T - h_{T-1})$. Moreover, the ansatz implies that $x_{T-1} = x_T$, as otherwise $Q(T-1, T) \neq 1$. Thus, the stochastic process reduces to the stochastic process of the $(p,c=1)$ case, with a trivial addition of $\delta(h_{T-1} - h_T) \delta_{x_{T-1}, x_T}$ for the last step. This implies that the order parameters up to time $T-1$ are the same as to the $(p,c=1)$ case, and all order parameters at time $T$ are trivially determined by the conditions $h_{T-1} = h_T, x_{T-1}= x_T$.

In practice, in the numerical solutions of the $(p,c=2)$ equations, we
solved both the $(p,c=1)$ and $(p,c=2)$ equations, and whenever we found two solutions with distinct properties, and in particular distinct entropy, we kept the solution with largest entropy as the dominant saddle-point. 

\subsection{Details for specific dynamical rules}
\label{app.dynamics}

In this paper we focus on a specific class of dynamical rules, which are subcases of the general rule
\begin{equation}
    \sigma(x, h) = 
    \sign((xh + \krl)(xh + \kgr)) \cdot x
    = 
    \begin{cases}
        - x &\mathif \kgr < - x h < \krl \\
        x   &\quad\text{otherwise} 
    \end{cases}\, ,
\end{equation}
where $ 0 \leq \kgr \leq \krl$ are fixed constants.
In this dynamics, a spin is flipped if it is misaligned with its local field, and its local field's magnitude satisfies $\kgr < |h| < \krl$ (this interpretation is valid only if $\kgr \geq 0$).

Some values of the thresholds $\kappa$ lead to special dynamical rules:
\begin{itemize}
    \item if $(\kgr, \krl) = (0, +\infty)$, the dynamics reduces to $\sigma(x, h) = \sign(h)$, the sync-Glauber dynamics.
    \item if $\kgr = \krl$, the dynamics is frozen $\sigma(x,h) = x$.
    \item if $\krl = +\infty$, the dynamics is given by $\sigma(x,h) = \sign(h + \kgr x)$, the sync-greedy dynamics. 
    \item if $\kgr = 0$, the dynamics is given by $\sigma(x,h) = \sign(xh^2 + \krl h)$, the sync-reluctant dynamics.
\end{itemize}

It is useful to provide the explicit form of \eqref{eq.fd-constr-del} for this class of dynamical rules.
The only non-trivial part is the value of the sets $I_{\pm}(x_t)$, which were defined as
$I_+(x_t)$, the set of points where the function $h_t \to \sigma(x_t, h_t)$ changes sign from $-1$ to $+1$, and  $I_-(x_t)$, the set of points where the function $h_t \to \sigma(x_t, h_t)$ changes sign from $+1$ to $-1$.
We have
\begin{equation}
    \begin{split}
        I_+(x=1) = \{ -\krl \} \, ,\quad
        I_-(x=1) = \{ -\kgr \} \, ,\quad
        I_+(x=-1) = \{ \kgr \} \, ,\quad
        I_-(x=-1) = \{ \krl \} \, .
    \end{split}
\end{equation}

\section{Numerical solutions of the DMFT equations}\label{app.numericsdmft}

In this section we provide some details on how to solve numerically the DMFT equations.

\subsection{Forward dynamics}

There are many strategies to solve the saddle-points equations for the forward DMFT \eqref{eq.fd-dmft}. We provide a brief summary below.

\subsubsection{Direct exact solution}

The equations are solved time-by-time, meaning that $Q(1,2), V(2,1)$ are computed at the beginning, then $Q(1,3), Q(2,3), V(3,1), V(3,2)$ are computed as a function of $\{Q(1,2), V(2,1)\}$, etc \dots up to the desired time $T$.
Each order parameter $Q(t,s), V(s,t)$ for $t<s$ is computed by numerically computing as precisely as possible \eqref{eq.fd-dmft}. 
We perform the sum over the $2^T$ trajectories $\ux$ exactly. 
For each trajectory $\ux$, the integral can be decomposed into a sum over Gaussian integrals over rectangular domains (as long as $h \to \sigma(x,h)$ changes sign only a finite number of times for each value of $x$).
Such integrals can be efficiently approximated by a quasi Monte Carlo algorithm \cite{genz}.

We call this solution scheme "direct" as it computes the order parameters time-by-time, and "exact", meaning that we numerically approximate  the summation/integration procedure.
This scheme becomes quickly computationally expensive with a complexity $\caO(T 2^T)$, due to a combination of exponential scaling of the summation and a linear scaling in the number of evaluation points for each integral.
In practice, we were able to solve the equations up to $T \approx 15$, depending on the specific choice of the parameters of the model.

\subsubsection{Direct sampling solution}

The equations are again solve time-by-time as above, but now the summation/integration is computed through a Monte Carlo approximation.
Indeed, as discussed in Section \ref{app.fd}, we can easily sample the stocastic process $p(\ux, \uh)$, and use it to approximate summation/integration by an empirical average.
For each new time step $T$ for which \eqref{eq.fd-dmft} needs to be solved, we generate $M$ samples from $p(\ux, \uh)$ and compute the correlators $Q(t, T) = \angavg{x_t x_T}$ and $\angavg{h_t x_T}$ for $1 \leq t < T$. Inverting \eqref{eq.Vxh} allows to compute $V(T, t)$.
This scheme is much more efficient than the "direct exact" scheme, and allows to easily reach $T \approx 30$ (and possibly larger, we did not study carefully the limits of this scheme). 

In the experiments for the model considered in this paper, this scheme provided a comparably good estimate of the order paramters with respect to the "direct exact" scheme, and was sensibly faster to run for times larger than $T > 4$. 
Thus, we decided to use this solution scheme to solve \eqref{eq.fd-dmft}.
We used at least $M = 10^6$ samples to esimtate the Monte Carlo integrals, leading to a relative error of order $\approx 10^-3$.

\subsubsection{Direct sampling solution with reuse}

The equations are solved as in the "direct sampling" scheme, but at each time step $T$ we save the trajectories used, and reuse them to generate the trajectories for time $T+1$ (as we discussed in Section \ref{app.fd}, the stochastic process at time $T+1$ can be sampled easily conditioned on its previous values $1 \leq t \leq T$).
This strategy was already discussed in detail \cite{opper94} in the specific case of $\sigma(x, h) = \sign(h)$.  

This scheme is computationally more efficient than the "direct sampling", but uses significantly more memory due to the necessity of storing the trajectories.
Also, as already noted in \cite{opper94}, the Monte Carlo approximation errors at different times will not be independent, and will compound as time grows.

In the experiments for the model considered in this paper, this scheme was not stable and precise enough compared to the "direct sampling" scheme, so we decided not to used it.

\subsubsection{Iterative solution}

The equations are not solved time-by-time. Instead, starting from an initial guess for the values of the order parameters, the equations (either computed as in the "direct exact" scheme, or as in the "direct sampling" scheme) are solved by fixed-point iteration.
This scheme has the advantage of avoiding the need to compute all the previous times $1 \leq t < T $ to get the order parameters at time $T$, as the equations are solved in one go. 
This is an actual advantage as long as the number of iterations required to reach convergence remains significantly smaller than $T$.

In the experiments for the model considered in this paper, this scheme was not stable and precise enough compared to the "direct sampling" scheme, so we decided not to used it.

\subsection{Backtracking dynamics}

For the case of the backtracking dynamics, the equations are non-causal, and the stochastic process is not easy to sample. For this reason, the only viable solutions scheme we found is the analogue of the "iterative solution" discussed above for the forward dynamics.
The equations's sums and integrals are estimated as in the "direct exact" method, by explicitly summing over all $2^T$ trajectories and by solving each Gaussian integral over a rectangular domain by quasi Monte Carlo integration \cite{genz} (with $5000 * T$ samples for $T$ total time).

As discussed above, this scheme becomes quickly computationally expensive with a complexity $\caO(T 2^T)$, due to a combination of exponential scaling of the summation and a linear scaling in the number of evaluation points for each integral. Moreover, there is the added complexity of having to iterate the equations multiple times until a fixed point is reached (we used an Anderson-accelerated iteration scheme from the NLsolve Julia package).

In practice, we were able to solve the equations up to $T = p+c \approx 11$, depending on the specific choice of the parameters of the model, up to a tolerance of $10^{-3}$ on the saddle point equations.

Finally, for the class of dynamics discussed in Section \ref{app.dynamics}, we solved the equations at fixed $(p,c)$ by finding a solution at a given choice of $(\kgr, \krl)$, and then using the already found solution as an initial condition for the solver for close values of the threshold parameters,
as it is usual the case for replica-style saddle-point equations.
For the sync-greedy dynamics, we usually solved both starting from $(0, +\infty)$ and increasing $\kgr$, and from $(+\infty, +\infty)$ and decreasing $\kgr$, and similarly for the sync-reluctant.

\end{document}